\documentclass[prb,twocolumn,aps,superscriptaddress]{revtex4-2}

\usepackage{graphicx}
\usepackage{dcolumn}
\usepackage{bm}
\usepackage{amssymb}
\usepackage{amsmath}
\usepackage{physics}
\usepackage{sublabel}
\usepackage[utf8]{inputenc}
\usepackage{hyperref}
\usepackage[usenames, dvipsnames]{color}
\usepackage[english]{babel}
\usepackage[T1]{fontenc}
\usepackage{bbm}

\usepackage{float}
\usepackage{verbatim}
\usepackage[caption=false]{subfig}
\usepackage{xfrac}
\usepackage{upgreek}
\hypersetup{colorlinks = true, linkcolor=blue, citecolor=blue, urlcolor=blue}

  \makeatletter
    \renewcommand\@make@capt@title[2]{%
     \@ifx@empty\float@link{\@firstofone}{\expandafter\href\expandafter{\float@link}}%
      {\textsc{#1}}\@caption@fignum@sep#2\quad}%
    \makeatother

\newcommand{\KP}{K^\prime}
\newcommand{\kup}{K\uparrow}
\newcommand{\kdown}{K\downarrow}
\newcommand{\kpup}{K^\prime\uparrow}
\newcommand{\kpdown}{K^\prime\downarrow}

\begin{document}


\title{Identifying Pauli blockade regimes in bilayer graphene double quantum dots}


\author{Ankan Mukherjee}
\affiliation{Department of Physics, Indian Institute of Technology Bombay, Powai, Mumbai-400076, India}
\author{Bhaskaran Muralidharan}
\affiliation{Department of Electrical Engineering, Indian Institute of Technology Bombay, Powai, Mumbai-400076, India}
\affiliation{Centre of Excellence in Quantum Information, Computation, Science and Technology, Indian Institute of Technology Bombay, Powai, Mumbai-400076, India}
\email{bm@ee.iitb.ac.in}


\date{\today}

\begin{abstract}
Recent experimental observations of current blockades in 2-D material quantum-dot platforms have opened new avenues for spin and valley-qubit processing. Motivated by experimental results, we construct a model capturing the delicate interplay of Coulomb interactions, inter-dot tunneling, Zeeman splittings, and intrinsic spin-orbit coupling in a double quantum dot structure to simulate the Pauli blockades. Analyzing the relevant Fock-subspaces of the generalized Hamiltonian, coupled with the density matrix master equation technique for transport across the setup, we identify the generic class of blockade mechanisms. Most importantly, and contrary to what is widely recognized, we show that conducting and blocking states responsible for the Pauli-blockades are a result of the coupled effect of all degrees of freedom and cannot be explained using the spin or the valley pseudo-spin only. We then numerically predict the regimes where Pauli blockades might occur, and, to this end, we verify our model against actual experimental data and propose that our model can be used to generate data sets for different values of parameters with the ultimate goal of training on a machine learning algorithm. Our work provides an enabling platform for a predictable theory-aided experimental realization of single-shot readout of the spin and valley states on DQDs based on 2D-material platforms.
\end{abstract}

\maketitle


\section{Introduction\label{sec:introdution}}

\begin{figure*}[thpb]
\centering
\captionsetup[subfigure]{oneside,margin={0.3cm,-1cm}}
\captionsetup[subfloat]{oneside,margin={0.3cm,-1cm},labelfont=bf}
%
%
\begin{minipage}{0.50\linewidth}
\centering
\hspace{-5mm}
    \subfloat[]{
    \begin{minipage}{0.60\linewidth}
        \includegraphics[width=\textwidth]{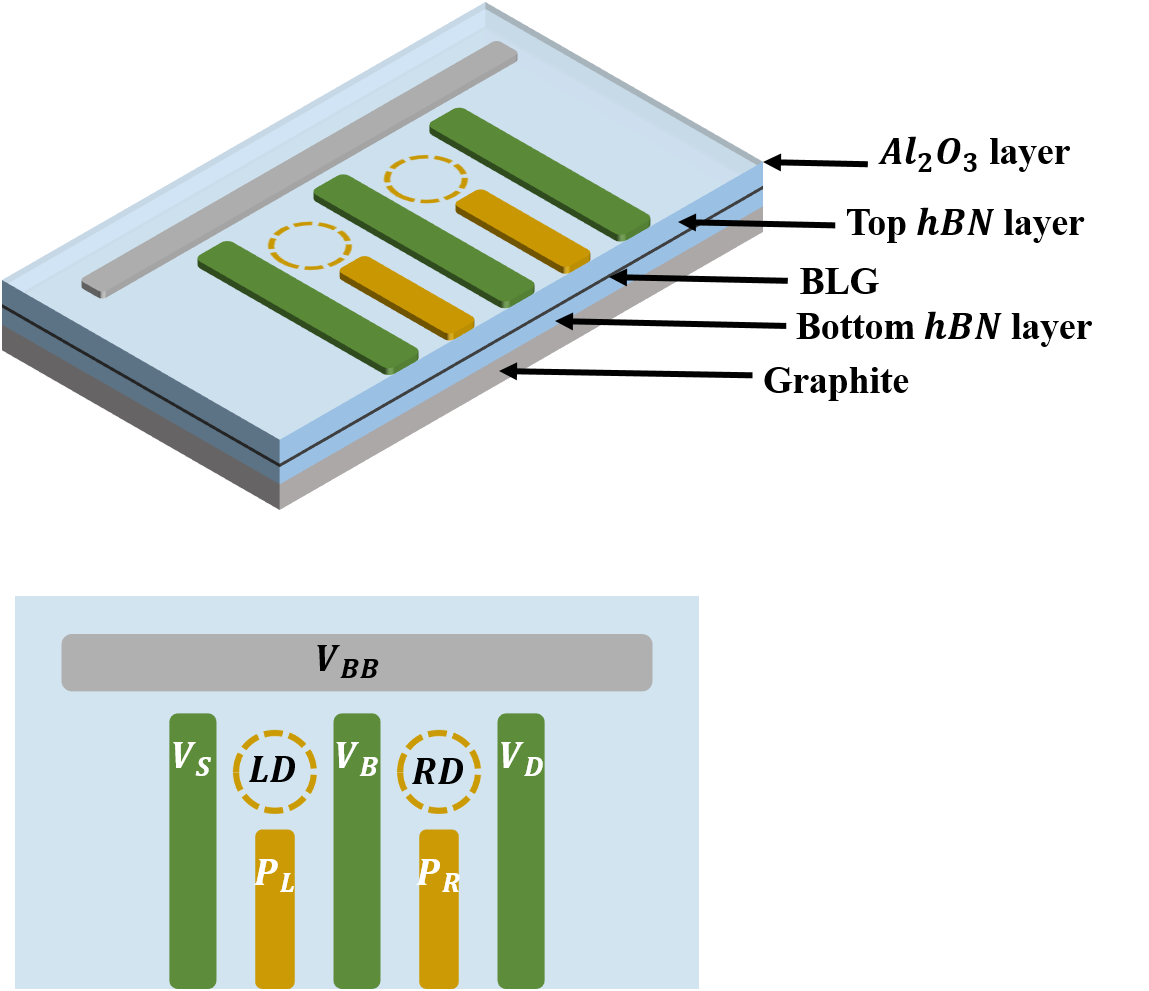}
    \end{minipage}}
    \subfloat[]{
    \begin{minipage}{0.40\linewidth}
        \includegraphics[width=1.05\textwidth]{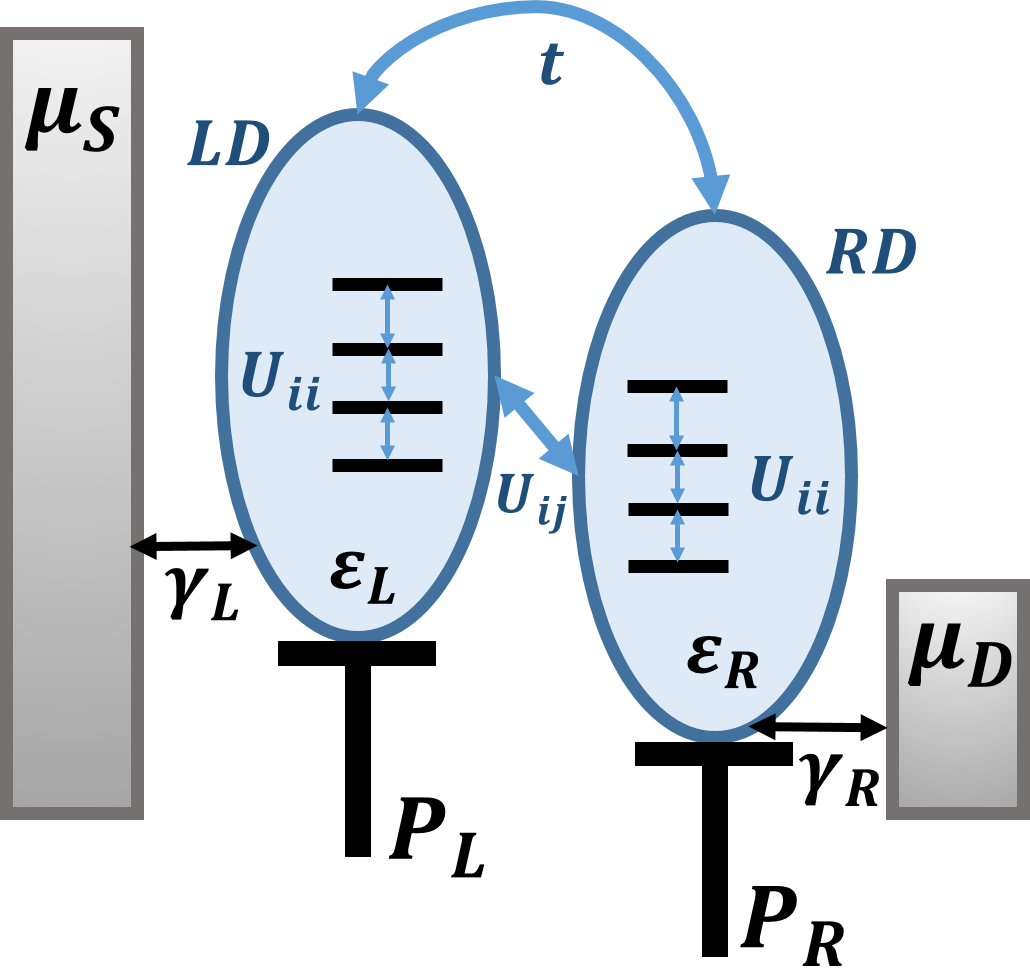}
    \end{minipage}}
    \\
    \hspace{-10mm}
    \subfloat[]{
    \begin{minipage}{\linewidth}
        \includegraphics[width=0.48\textwidth]{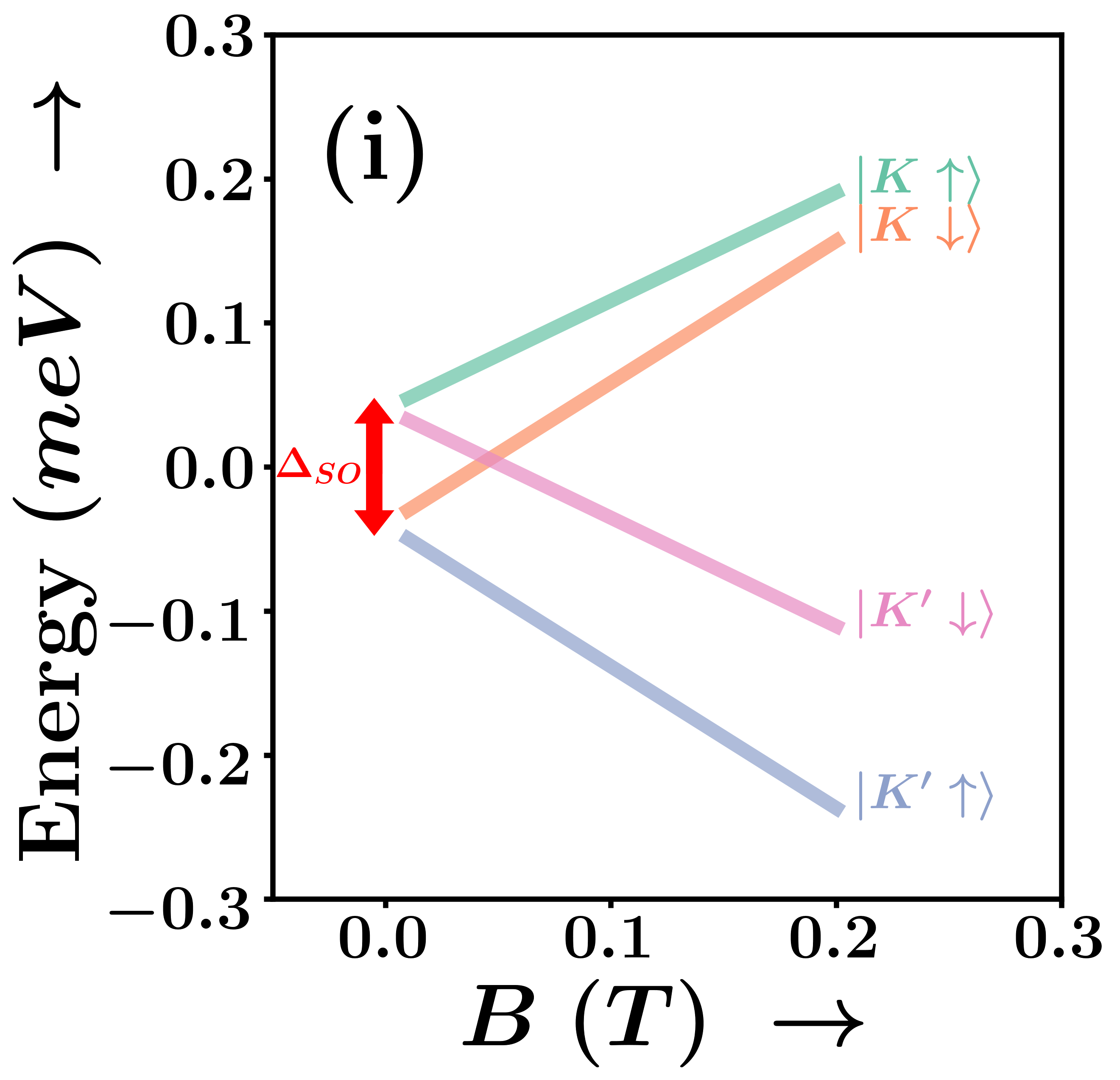}\hspace{2mm}
        \includegraphics[width=0.48\textwidth]{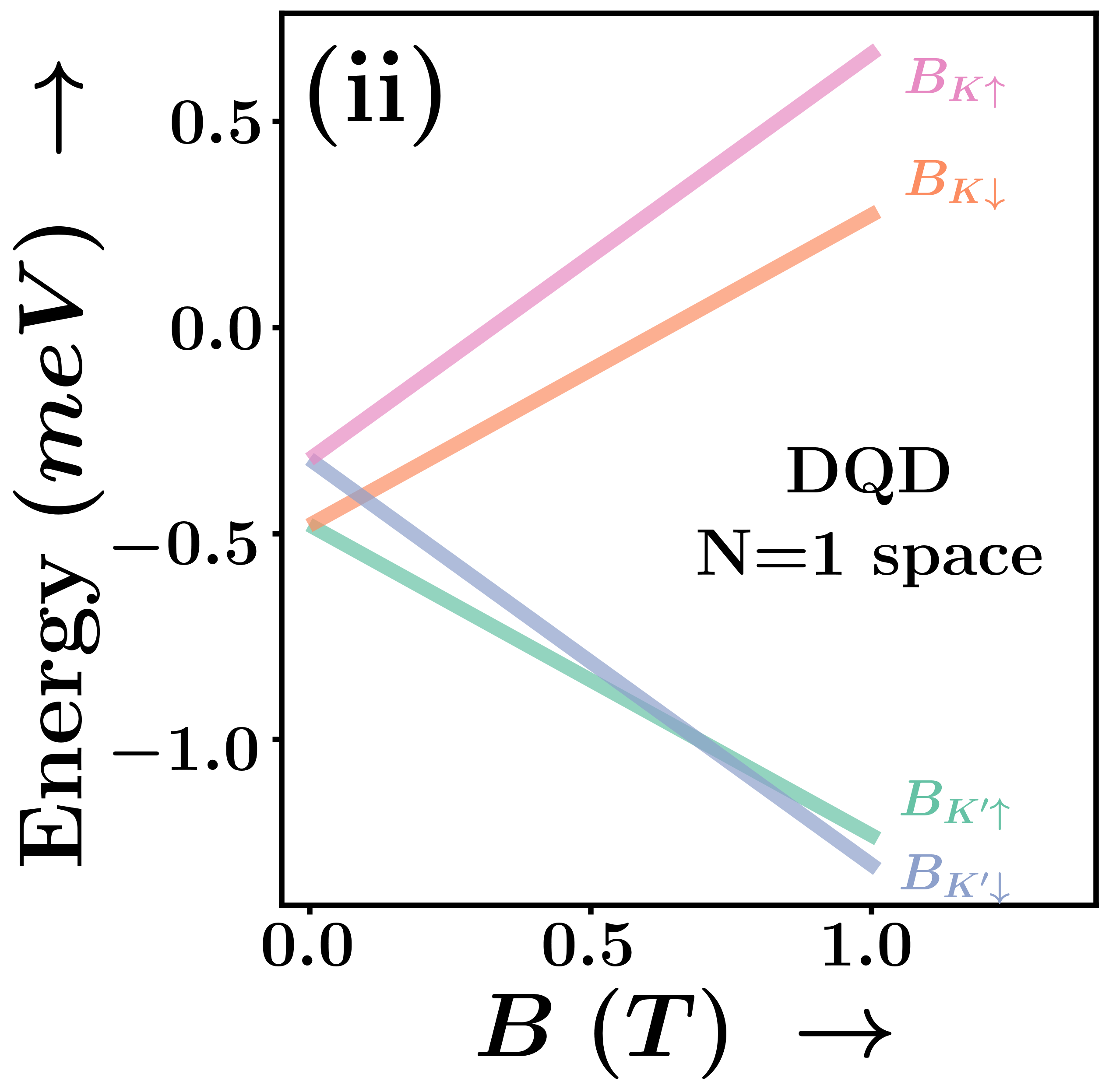}\vspace{-3mm}
    \end{minipage}}
\end{minipage}\hspace{5mm}
\begin{minipage}{0.38\linewidth}
    \subfloat[]{
    \begin{minipage}{\linewidth}
        \includegraphics[width=\textwidth]{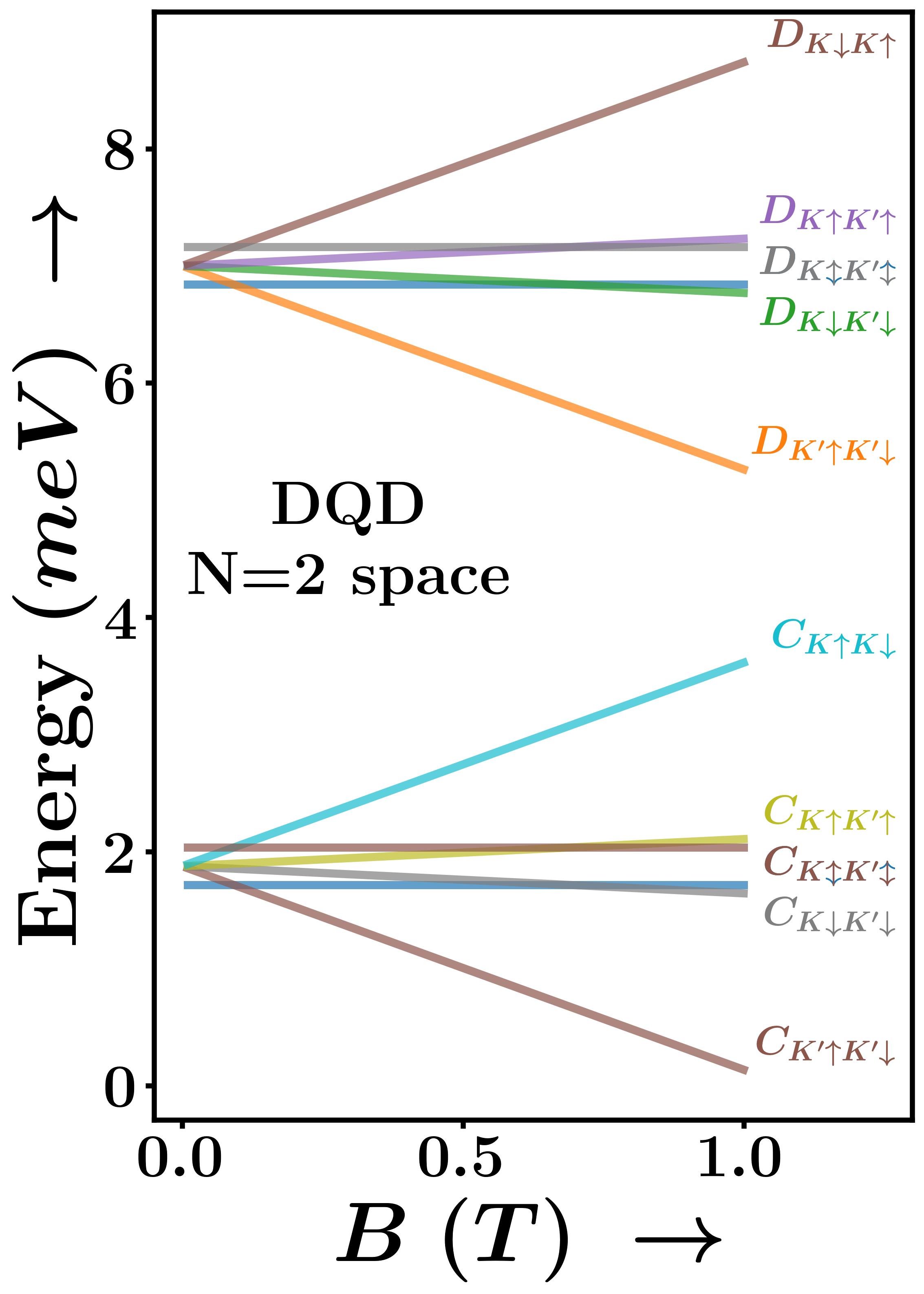}
    \end{minipage}}
\end{minipage}
%
%
\caption{Device schematics and the relevant Fock-subspaces. 
\textbf{(a)} Pseudo-color schematic of the device used. The potentials $V_S$ and $V_D$ control the source-drain bias, $V_B$ is the barrier voltage used to control the tunneling parameter $t$. $P_L$ and $P_R$ are the plunger voltages that determine the onsite energy of the left dot ($LD$) ($\varepsilon_L$) and right dot ($RD$) ($\varepsilon_R$) respectively.
\textbf{(b)} Schematic of the energy states in each dot. The parameters are demonstrated, and the meaning of each symbol is described in the main text. Each dot bears four states arising from the spin and valley DOFs.
\textbf{(c)} Energy diagrams as a function of magnetic field: (i) shows the energy states of a single dot for a single electron occupancy. The energy diagram is a consequence of intrinsic spin-orbit coupling and spin and valley Zeeman effects; (ii) shows the energies of the DQD bonding states for the $N=1$ occupancy subspace.
\textbf{(d)} Relevant Fock space energies for the $N=2$ conducting and blocking (dark) states. The meaning of the notation used has been described in the main text. The $C$ states shown are the ones with the lowest energies.
}
\label{fig:intro_fig}
\end{figure*}

Semiconductor spin-qubits on Si and Ge material platform \cite{spin_qubit1,spin_qubit2,spin_qubit3} have been strongly pursued for spin-based quantum computing and quantum information processing \cite{liu,brooks,sala,lei} over the past decade, due to the ease of measurement and readout, and long spin-coherence times \cite{longer_coherence_and_fidelity,line_shapes_pauli_blockade}. Two dimensional (2-D) materials such as bi-layer graphene (BLG) and transition metal di-chalcogenides (TMDC) \cite{liu,Giustino2020,2d_roadmap1,graphene_2D,spin_qubits_in_graphene,graphene_single_shot_readout,graphene_coherence,2d_hyperfine3}, owing to a negligible hyper-fine interaction \cite{2d_hyperfine1,2d_hyperfine2,2d_hyperfine3,TMDC_paper}, have sparked a lot of current interest for better control over initialization and readout of spin-qubits. The non-vanishing Berry curvature at the degeneracy points \cite{blg_berry_curvature1,blg_berry_curvature2}, creates an additional valley degree of freedom that couples with the spin and gives additional venues for information processing. \\
\indent The initialization and readout of spin-qubits typically involves the Pauli spin-blockade (PSB) in electronic transport \cite{single_shot_original,spin_filtering_petta,first_spin_qubit}, whose mechanism is a spin-selection rule \cite{bhasky_main} between conducting and blocking states. In 2-D materials, the valley degree of freedom creates multiple pathways between conducting and blocking states, leading to a general class of Pauli blockades, intertwining the spin and valley degrees of freedom. While PSB has been studied extensively on Si \cite{present_model4,si_lai,si_tadokoro} and Ge platforms \cite{ge_georgios,ge_wang}, and given several advances in experiments across 2-D platforms, \cite{so_exp_blg,tunable_interdot_blg,tunable_bandgap_blg,insulating_state_in_blg,spin_and_valley_blg,charge_sensing_blg,valley_only_qubit_blg, main_exp_blg_2, main_exp_blg}, understanding the general class of Pauli blockades across 2-D material quantum dot platforms remains an unexplored aspect. For instance, recent experiments on carbon nanotubes \cite{cnt_1,cnt_2,cnt_3,cnt_4}, BLGs \cite{graphene_2D,graphene_coherence,graphene_single_shot_readout}, quantum wells \cite{brooks,lei,sala}, and TMDCs \cite{tmdc_qubits1,tmdc_qubits2} have confirmed that the valley pseudo-spin creates a blockade even when spin-blockade is absent. \\ 
\indent In this paper, we identify the causes of the different Pauli blockades, build a model for the transport mechanism through the quantum dots, and build a generic theory for predicting blockade regimes on 2-D materials. Building on a double quantum dot (DQD) transport setup \cite{dqd_structure1,dqd_structure2,BM_1,Muralidharan_2008}, schematized in Fig.~\ref{fig:intro_fig}(a), we perform the analysis of Pauli blockades for the regime where the total occupancy of the dots is 2 electrons, since this is the most commonly studied regime in experimental literature \cite{main_exp_blg,dqd_2e}. Our models take into account \cite{hubbard1,hubbard2,hubbard3,TMDC_paper}, intrinsic spin-orbit (SO) coupling, spin-Zeeman splitting, valley-Zeeman splitting, a weak inter-dot tunneling that preserves spin and valley pseudo-spin, and Coulomb interactions, as schematized in Fig.~\ref{fig:intro_fig}(b). \\ 
\indent Analyzing the relevant Fock-subspaces of the generalized Hamiltonian, coupled with the density matrix master equation technique for transport across the setup, we identify the generic blockade mechanisms. Most importantly, and contrary to what is widely recognized \cite{main_exp_blg,main_exp_blg_2}, we show that conducting and blocking states responsible for the Pauli-blockades are a result of the coupled effect of all degrees of freedom and cannot be explained using the spin or the valley pseudo-spin on its own. We then numerically predict the regimes where Pauli blockades might occur. To this end, we verify our model against actual experimental data \cite{main_exp_blg}, and propose that our model can be used to generate data sets for different values of parameters with the ultimate goal of training on a machine learning algorithm. Our work, we believe, provides an enabling platform for a predictable theory-aided experimental realization of single-shot readout of the spin as well as valley states on DQDs based on 2D-material platforms.\\
\indent The paper is organized as follows. In Sec.~\ref{sec:formalism}, we construct the effective Hamiltonian for the DQD, solve for the relevant eigenstates, and obtain the equations governing the flow of current. We present the conditions under which current blockade may be realized and introduce the terminologies "conducting" and "dark" to classify the eigenstates based on their contribution towards the current. In Sec.~\ref{sec:results}, we perform simulations on our model. We begin with the study of the effect of varying the source-drain bias on the current and correlate blockade regimes with the state transitions. We then vary the onsite energies to obtain the charge-stability diagram for a DQD and locate the bias triangles for three different values of the external magnetic field. We also observe the behavior of the current in these bias triangles and justify the occurrence of multiple blockades. Finally, we present our conclusions and scope for future work in Sec.~\ref{sec:conclusion}.


\section{Formalism and Methods\label{sec:formalism}}
We start with a 2-D material platform on which a DQD is created by virtue of confinement using voltage-controlled gates \cite{DQD1,DQD2}, as demonstrated in Fig.~\ref{fig:intro_fig}(a). We have a left(right) quantum dot $LD(RD)$, whose onsite energy, $\varepsilon_L(\varepsilon_R)$ can be controlled using the lever arm voltage $P_L(P_R)$. We assume the cross-talk between the gates $P_L$ and $P_R$ to be zero through the construction of appropriate virtual gates \cite{virtual_gates1,virtual_gates2}. The model abstraction for the DQD system is illustrated in Fig.~\ref{fig:intro_fig}(b). The source drain bias voltage $V_{SD}$ is applied across the electrodes labelled $V_S$ and $V_D$. The voltage at gate $V_B$ controls $t$, the inter-dot tunneling, which, in our model, preserves the spin and the valley pseudo-spin. The backbone voltage $V_{BB}$ can be tuned to control the overall degree of confinement.

\subsection{DQD model\label{subsec:formalism:model}}
We model the system using a modified Hubbard Hamiltonian \cite{qtt,TMDC_paper}. Besides the onsite energies and the inter-dot tunneling, there is an onsite Coulomb repulsion, with energy $U_{ii}$, between each pair of electrons on each of the dots, and an inter-dot Coulomb repulsion with energy $U_{ij}$. In principle, $U_{ii}$ and $U_{ij}$ can be matrices indexed by $i$ and $j$. On each dot, the conduction electrons are localized on either of the two valleys: $K$ or $\KP$, and can have a spin $\uparrow$ or spin $\downarrow$. Thus, for a single dot, there are $4$ states available, namely, $\ket{\kup},\ket{\kdown},\ket{\kpup},\ket{\kpdown}$. We shall henceforth use the symbol $\zeta$ to index these four states. In the absence of an external magnetic field, the four energy states are split into two Kramer pairs by an intrinsic spin-orbit (SO) coupling \cite{so_theory_blg1,so_theory_blg2,so_exp_blg,so_exp1,so_exp2}, as shown in Fig.~\ref{fig:intro_fig}(a). The energy of the pair $\{\ket{\kup},\ket{\kpdown}\}$ is increased by an amount $\frac{1}{2}\Delta_{SO}$, while that of the pair $\{\ket{\kdown},\ket{\kpup}\}$ is decreased by the same amount. In the presence of an external magnetic field, the degeneracy of the states in each of the Kramer pairs is broken by the spin-Zeeman and the valley-Zeeman effects. The energy shift due to the spin-Zeeman splitting is given as $h_S=\sigma g_s\mu_BB$, while that due to the valley-Zeeman splitting is given by $h_V=\tau g_v\mu_BB$, where, $\sigma$ is the spin ($\sigma=+\frac{1}{2}$ for spin $\uparrow$, and $\sigma=-\frac{1}{2}$ for spin $\downarrow$) and $\tau$ is the valley pseudo-spin ($\tau=+\frac{1}{2}$ for valley $K$, and $\tau=-\frac{1}{2}$ for valley $\KP$). $\mu_B=5.79\times10^{-5}$ eVT$^{-1}$ is the electron magnetic moment, $B$ is the external magnetic field applied perpendicular to the plane of the BLG, and $g_s$ and $g_v$ are the spin and valley g-factors respectively. Under such considerations, the Hamiltonian takes the form
\begingroup
\allowdisplaybreaks
\begin{align}
    \hat{H}_{DQD} &= \underbrace{\varepsilon_L \hat{n}_L + \varepsilon_R \hat{n}_R}_{\text{Onsite energy}}\nonumber\\
    &\hspace{2mm}+\underbrace{\frac{U_{ii}}{2}\left(\hat{n}_L^2-\hat{n}_L+\hat{n}_R^2-\hat{n}_R\right)}_{\text{Onsite repulsion}}%
    +\underbrace{U_{ij}\hat{n}_L\hat{n}_R}_{\text{Interdot replusion}}\nonumber\\
    &\hspace{4mm}+\underbrace{\sum_{\tau,\sigma}t\hat{c}_{R\sigma\tau}^\dagger \hat{c}_{L\sigma\tau} + \text{h.c.}}_{\text{Interdot tunneling}}\nonumber\\
    &\hspace{6mm}+\underbrace{\frac{\Delta_{SO}}{2}\sum_{\alpha,\tau,\sigma}\hat{c}_{\alpha\sigma\tau}^\dagger\left(\bm{\sigma}_3\right)_{\sigma\sigma}\left(\bm{\tau}_3\right)_{\tau\tau}\hat{c}_{\alpha\sigma\tau}}_{\text{Spin orbit coupling}}\nonumber\\
    &\hspace{8mm}+\underbrace{\left|h_S\right|\sum_{\alpha,\tau,\sigma}\hat{c}_{\alpha\sigma\tau}^\dagger\left(\bm{\sigma}_3\right)_{\sigma\sigma}\hat{c}_{\alpha\sigma\tau}}_{\text{Spin Zeeman effect}}\nonumber\\
    &\hspace{8mm}+\underbrace{\left|h_V\right|\sum_{\alpha,\tau,\sigma}\hat{c}_{\alpha\sigma\tau}^\dagger\left(\bm{\tau}_3\right)_{\tau\tau}\hat{c}_{\alpha\sigma\tau}}_{\text{Valley Zeeman effect}}\label{eq:fermi_hubbard_hamiltonian},
\end{align}%
\endgroup
where the summations are defined over $\alpha\in\{L,R\}$, $\sigma\in\{\uparrow,\downarrow\}$, $\tau\in\{K,\KP\}$. The terms $\bm{\sigma}_3$ ($\bm{\tau}_3$) is the z-component of the Pauli matrix for the spin (valley pseudo-spin), defined as 
\begin{subequations}
\begin{alignat}{4}
    \left(\bm{\sigma}_3\right)_{\uparrow\uparrow}&=1;\quad\left(\bm{\sigma}_3\right)_{\uparrow\downarrow}&&=\left(\bm{\sigma}_3\right)_{\downarrow\uparrow}&&=0;\quad\left(\bm{\sigma}_3\right)_{\downarrow\downarrow}&&=-1\\
    \left(\bm{\tau}_3\right)_{KK}&=1;\quad\left(\bm{\tau}_3\right)_{K\KP}&&=\left(\bm{\tau}_3\right)_{\KP K}&&=0;\quad\left(\bm{\tau}_3\right)_{\KP\KP}&&=-1
\end{alignat}
\end{subequations}
The symbol $\hat{n}_L$ ($\hat{n}_R$) denotes the number operator for the number of electrons on the left (right) dot, formulated as
\begin{align}
    \hat{n}_\alpha=\sum_{\sigma,\tau}\hat{c}^\dagger_{\alpha\sigma\tau}\hat{c}_{\alpha\sigma\tau},
\end{align}
where $\hat{c}^{(\dagger)}_{\alpha\sigma\tau}$ is the annihilation (creation) operator for an electron on dot $\alpha$ with spin $\sigma$ in valley $\tau$. Fig.~\ref{fig:intro_fig}(c)(i) shows the energy of the four basis states as a function of the external magnetic field $B$.

The above Hamiltonian takes a block diagonal form and separates into nine sub-spaces, each with an invariant total number of electrons $N=0,1,\cdots,8$. In discussing the blockades in the two electron occupancy regime, only the subspaces $N=1$ and $N=2$ are relevant.
We use $L_\zeta\left(R_\zeta\right)$ to denote that there is an electron in the state $\zeta$ on the left(right) quantum dot, where $\zeta\in\{\kup, \kdown, \kpup, \kpdown\}$. For instance, a system with two electrons: one in the left dot in state $\kup$ and the other in the right dot in state $\kpdown$ is represented as $\ket{L_{\kup} R_{\kpdown}}$. We also develop the notation $(n_L,n_R)$ to represent a state with $n_L$ electrons on the left dot and $n_R$ electrons on the right dot. The eigenstates are not the $(n_L,n_R)$ states, but a superposition of $(n_L,n_R)$ states with $n_L+n_R=N=\text{constant}$.

\subsection{Fock subspaces of the Hamiltonian\label{subsec:formalism:fock}}
The $N=1$ sub-matrix of Hamiltonian~\eqref{eq:fermi_hubbard_hamiltonian} is an $8\times8$ matrix and thus has eight eigenstates. We classify them into two groups of states, the bonding, and the anti-bonding states.
\begin{subequations}
\begin{alignat}{2}
\ket{B_{\zeta}} &= \xi\ket{L_\zeta}+\eta\ket{R_\zeta}\\
\ket{AB_{\zeta}} &= \xi\ket{L_\zeta}-\eta\ket{R_\zeta}.
\end{alignat}
\label{eq:N_1_fockspace}
\end{subequations}
In the above formulation, $B$ represents the bonding states and are lower in energy, while $AB$ represents the anti-bonding states and are higher in energy. $\zeta\in\{\kup, \kdown, \kpup, \kpdown\}$. Thus, we have four bonding and four anti-bonding states. Fig.~\ref{fig:intro_fig}(c)(ii) shows the energies of the bonding eigenstates as a function of the magnetic field applied.

The $N=2$ sub-matrix of the Hamiltonian is a $28\times28$ matrix with twenty eight eigenstates. Under the approximations of very weak inter-dot tunneling, small detuning, and zero inter-dot Coulomb repulsion, the eigenstates can be represented as products of spin and valley singlets and triplets \cite{TMDC_paper,main_exp_blg_2,main_exp_blg}. In this paper, we solve the Hamiltonian completely without any approximation. We then categorize the eigenstates, not according to their spins or valley pseudo-spins, but according to their contribution towards the current through the DQD. The eigenstates can be classified into three broad categories as follows.
\begin{subequations}
\begin{alignat}{2}
\ket{C_{\zeta_1\zeta_2}} &= \alpha\left(\ket{L_{\zeta_1}R_{\zeta_2}}-\ket{L_{\zeta_2}R_{\zeta_1}}\right)\nonumber\\
&\hspace{20mm}+\beta\ket{L_{\zeta_1}L_{\zeta_2}}+\kappa\ket{R_{\zeta_1}R_{\zeta_2}}\\
\ket{D_{\zeta_1\zeta_2}} &= \frac{1}{\sqrt{2}}\left(\ket{L_{\zeta_1}R_{\zeta_2}}+\ket{L_{\zeta_2}R_{\zeta_1}}\right)\\
\ket{P_{\zeta}} &= \ket{L_{\zeta}R_{\zeta}},
\end{alignat}
\label{eq:N_2_fockspace}
\end{subequations}
where $\zeta\in\{\kup, \kdown, \kpup, \kpdown\}$ and the combination $\zeta_1\zeta_2\in\{\kup\kdown,\ \kup\kpup,\ \kup\kpdown,\ \kdown\kpup,\ \kdown\kpdown,\ \kpup\kpdown\}$. $\zeta_1$ and $\zeta_2$ are taken over all possible unordered combinations of $\zeta$. We, therefore, have six possible combinations of $\zeta_1\zeta_2$. Each of the states $C$ occurs threefold, with three different sets of values of $\alpha$, $\beta$, and $\kappa$. Thus, in total, we have eighteen $C$ states, six $D$ states, and four $P$ states. The state labels $C$ and $D$ stand for "conducting" and "dark" respectively, according to their role in the equation for current, as is explained in Sec.~\ref{subsec:formalism:current_collapse}. The energies of the states in each of the three sets of the $C$ states differ significantly, allowing us to choose only the first unique set of lowest energy states in $C$ to model the current. Figure~\ref{fig:intro_fig}(d) shows the energies of the $D$ and the lowest energy $C$ eigenstates as a function of the magnetic field applied. We provide a detailed discussion of the eigenstates and their properties in appendix~\ref{sec:app1}. We proceed to build the framework for the transport mechanism through the dots.

\begin{figure*}[thpb]
\centering
\captionsetup[subfigure]{oneside,margin={0.3cm,-1cm}}
\captionsetup[subfloat]{oneside,margin={0.3cm,-1cm},labelfont=bf}
\hspace{-8mm}
\subfloat[]{
\begin{minipage}{0.32\linewidth}
    \centering
    \includegraphics[width=0.65\linewidth]{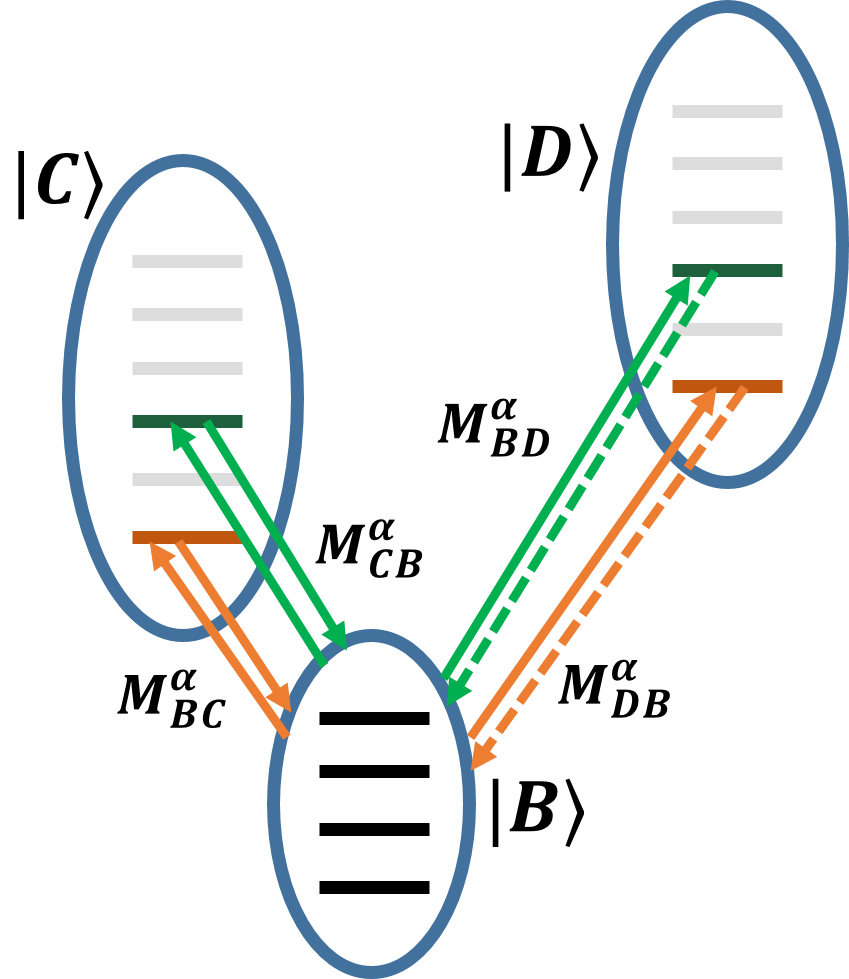}
\end{minipage}}\hspace{-8mm}
\subfloat[]{
\begin{minipage}{0.35\linewidth}
    \centering
    \hspace{-8mm}\includegraphics[width=0.65\linewidth]{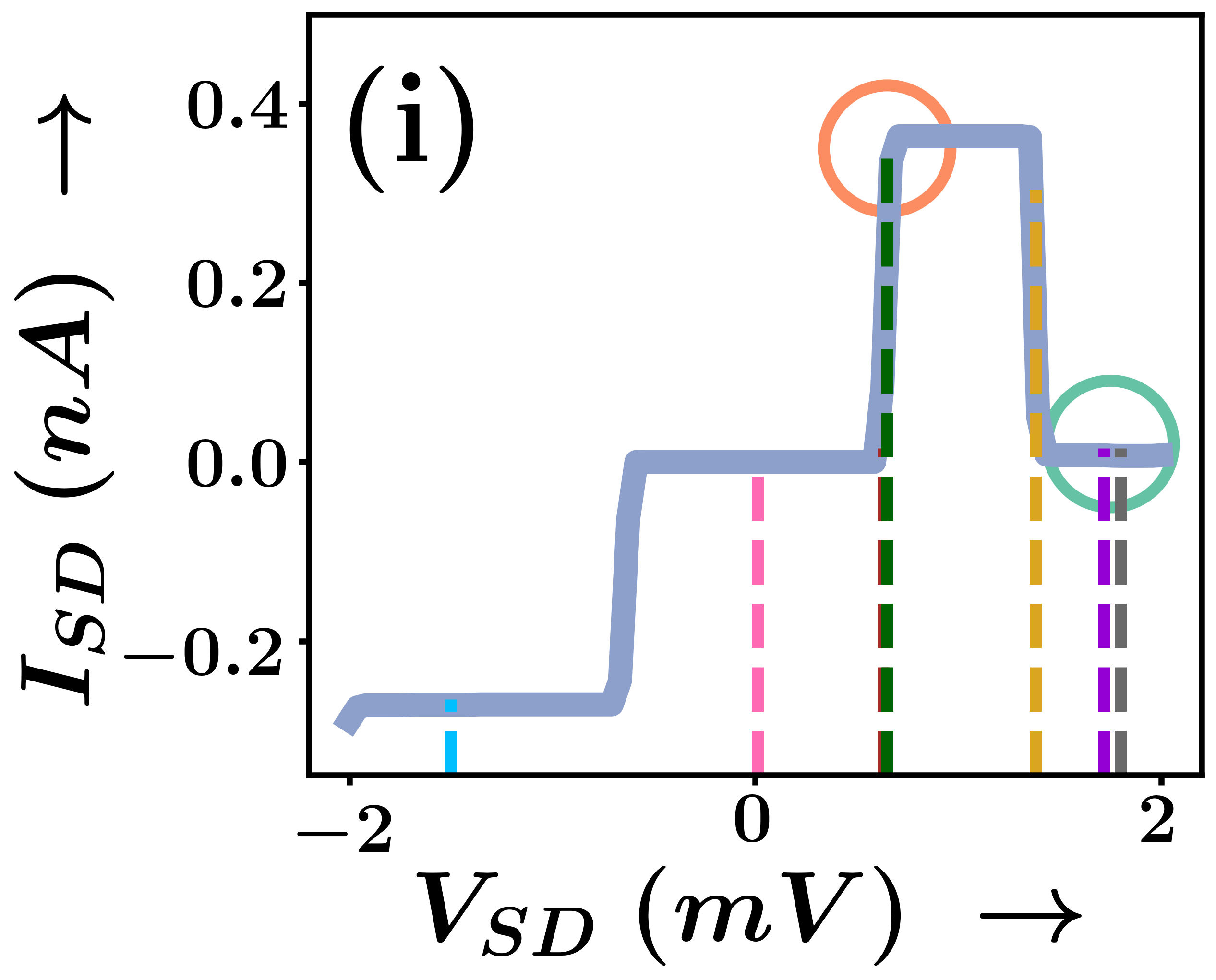}
    \begin{minipage}{\linewidth}
        \centering
        \includegraphics[width=0.48\linewidth]{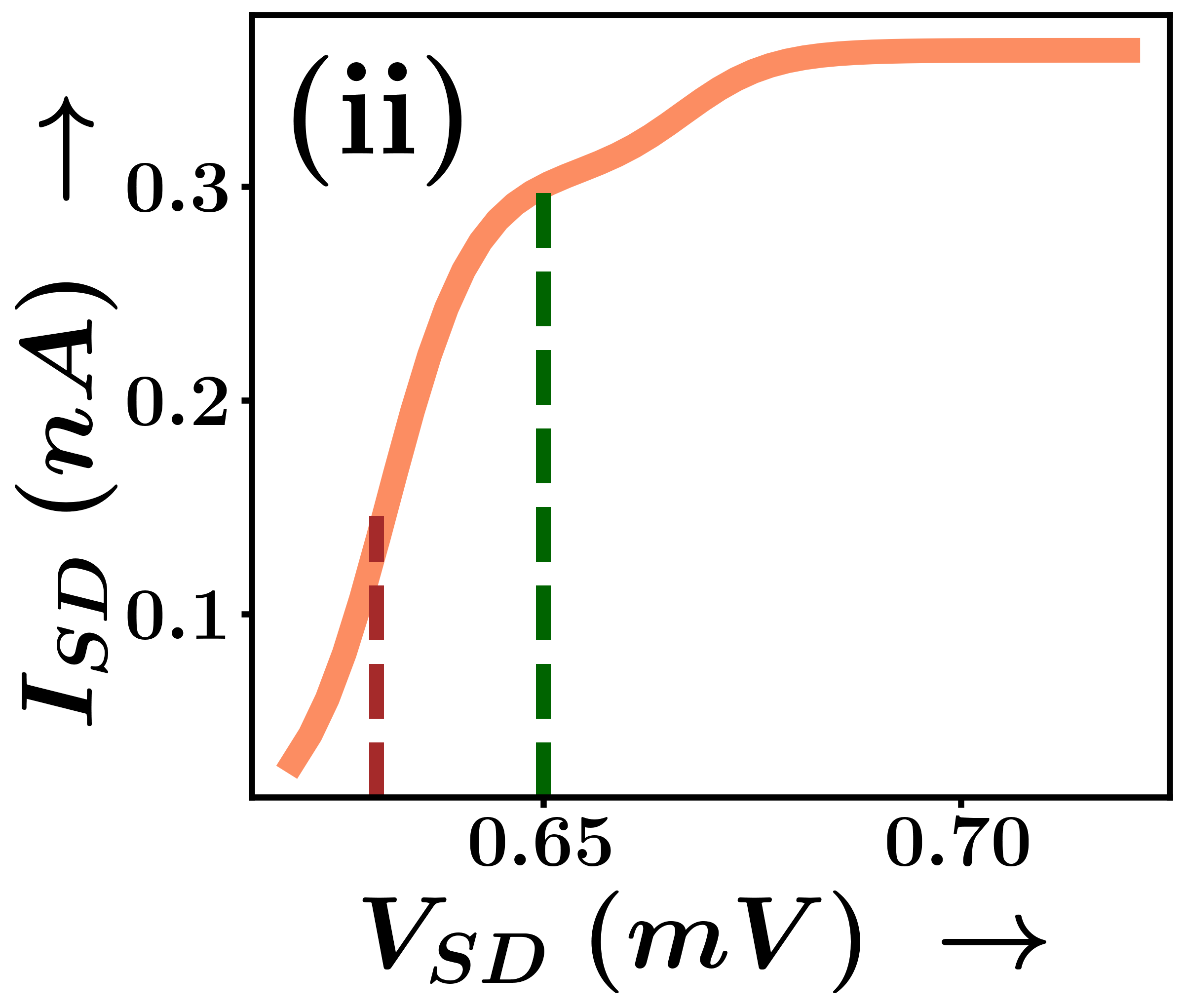}
        \includegraphics[width=0.48\linewidth]{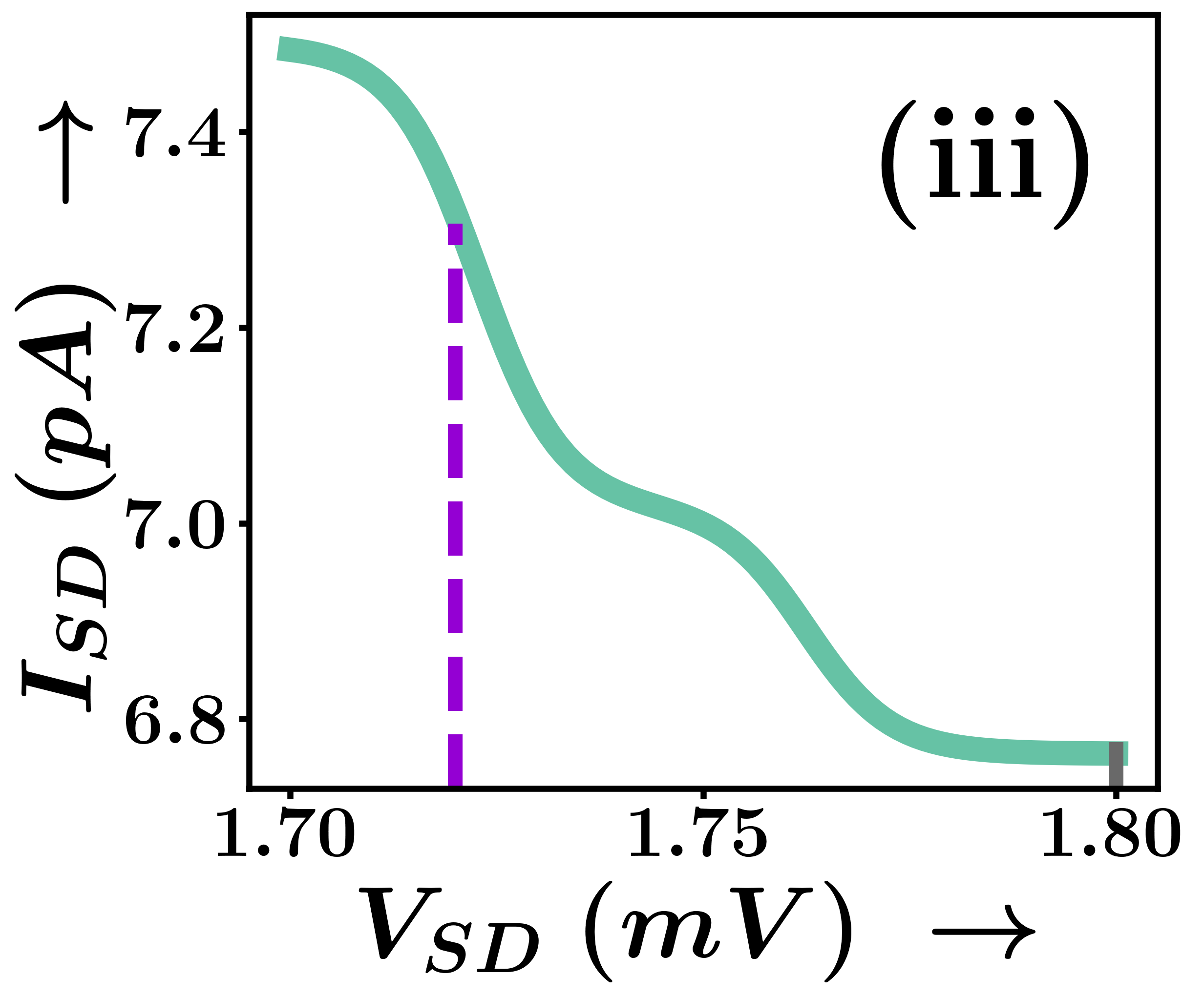}
    \end{minipage}
\end{minipage}}
\subfloat[]{
\begin{minipage}{0.35\linewidth}
    \includegraphics[width=\textwidth]{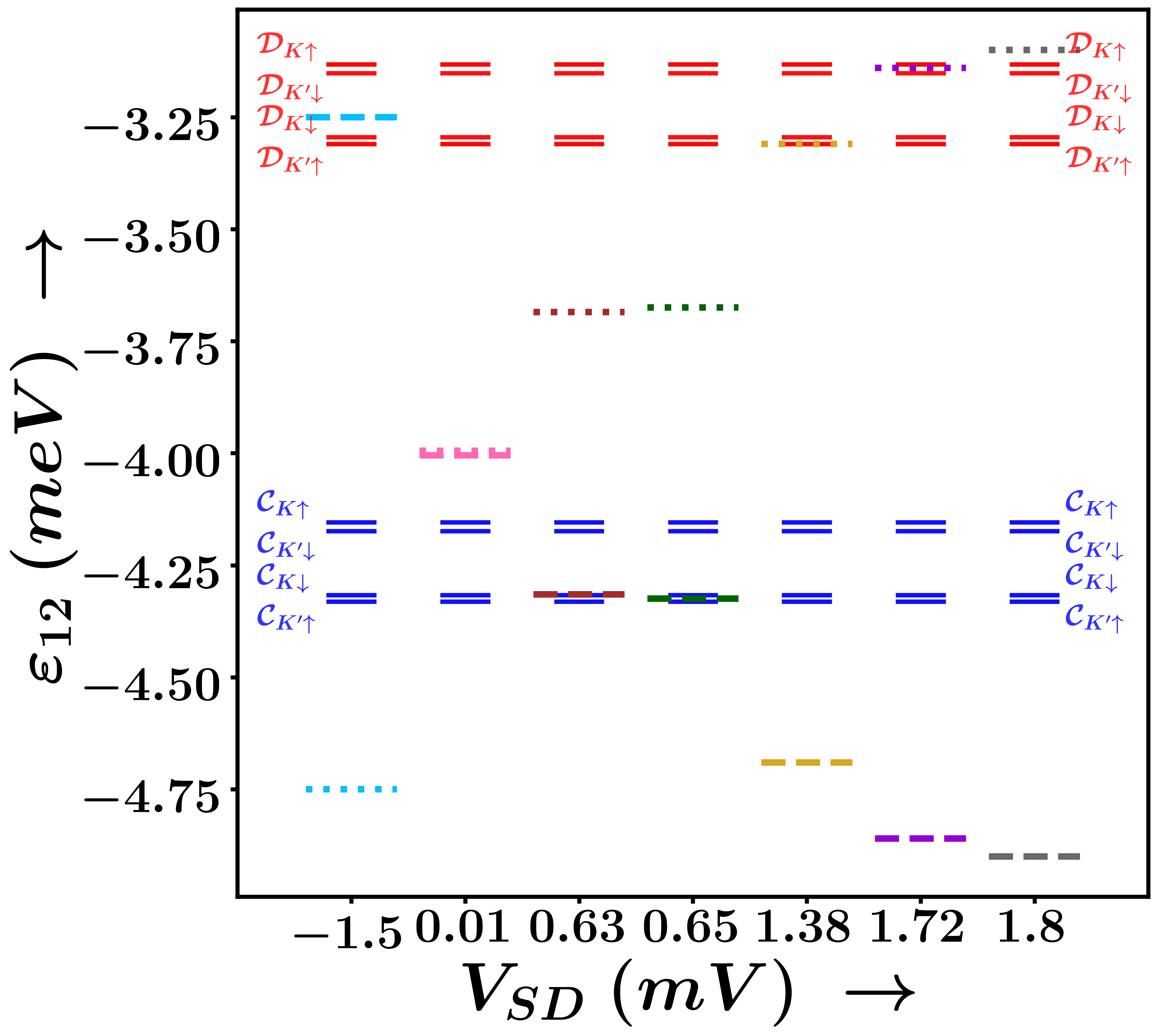}
\end{minipage}}
\captionsetup{justification=raggedright,singlelinecheck=false}
\caption{
Occurrence of current collapse leading to Pauli blockade for $\varepsilon_L=-5.25$ meV, $\varepsilon_R=-10.95$ meV, $B=0.01$ T. A low temperature of $30$ mK has been chosen to obtain distinct plateaus.
\textbf{(a)} Energy states and two of the allowed transitions from the $N=1$ to the $N=2$ subspace. The bold arrows indicate conducting transitions, while the dotted arrows indicate dark transitions. Access to the dark transitions in presence of the conducting transitions leads to the current blockade.
\textbf{(b)} Current through the dots as a function of the source-drain bias, $V_{SD}=V_S-V_D$. The orange and green regions in (i) are zoomed in (ii) and (iii), respectively.
\textbf{(c)} Transition energies and the chemical potentials of the source and the drain at each of the colored vertical line cuts in (b). The blue lines indicate the transitions $\mathcal{C}$, while the red lines indicate the transitions $\mathcal{D}$.
The blockade mechanism is explained in section~\ref{subsec:results:blockade}.
}
\label{fig:ndr_physics}
\end{figure*}

\subsection{Transport formulation\label{subsec:formalism:transport}}
While there is a strong understanding of the theory of current blockades in DQDs with a single (spin) degree of freedom \cite{bhasky_main,bhasky_2,ono,benzene,dark_channels}, the current through DQDs with multiples degrees remains an untackled challenge. The total current in the system results from a complex interplay of the probability of occupation of each eigenstate and the rates of transition between them. To tackle this problem, we extend the master equation prevalent in literature \cite{bhasky_2} to our model. The eigenstates are labelled $\ket{N,i}$, where $N$ denotes the total electron occupancy and $i$ denotes the $i^\text{th}$ state in the corresponding Fock state subspace with total electron occupancy $N$. We define the quantity $P^N_i$ to denote the probability of occupancy of the state $\ket{N,i}$ and $R^{L(R)}_{(N_1,i)\rightarrow(N_2,j)}$ to denote the rate of transition from the state $\ket{N_1,i}$ to the state $\ket{N_2,j}$ by virtue of injection or removal of an electron from the source(drain). Henceforth, we shall use the index $\alpha$ for the source ($\alpha=L$) or the drain ($\alpha=R$). The probabilities $P^N_i$ evolve over time as
\begin{align}
    \dot{P}^N_i &= \sum_j \left[R_{(N\pm1,j)\rightarrow(N,i)}P^{N\pm1}_j-R_{(N,i)\rightarrow(N\pm1,j)}P^N_i\right],\label{eq:master_equation}
\end{align}
where 
\begin{align}
    R_{(N_1,i)\rightarrow(N_2,j)}=\sum_{\alpha\in\{L,R\}}R^{\alpha}_{(N_1,i)\rightarrow(N_2,j)},
\end{align}
The rates depend on the transition matrix elements. We consider transport in the first order so that terms are non-zero if and only if $|N_1-N_2|=1$. We express the rates in terms of the matrix elements as
\begin{subequations}
\begin{alignat}{2}
    R^{\alpha}_{(N,i)\rightarrow(N-1,j)}&=\Gamma^{Nr}_{{\alpha} ij}\left[1-f\left(\frac{\epsilon^{Nr}_{ij}-\mu_{\alpha}}{k_BT}\right)\right]\\
    R^{\alpha}_{(N,i)\rightarrow(N+1,j)}&=\Gamma^{Na}_{{\alpha} ij}f\left(\frac{\epsilon^{Na}_{ij}-\mu_{\alpha}}{k_BT}\right),
\end{alignat}
\end{subequations}
where
\begin{subequations}
\begin{alignat}{2}
    \epsilon^{Nr}_{ij}&=E^N_i-E^{N-1}_j\\
    \epsilon^{Na}_{ij}&=E^{N+1}_i-E^N_j,
\end{alignat}
\end{subequations}
where $E^N_i$ is the eigenenergy of the state $\ket{N,i}$, $k_B$ is the Boltzmann's constant, $T$ is the temperature of the system (we assume this to be uniform across the entire system), $f$ is the Fermi-Dirac distribution, and $\Gamma^{Nr(a)}_{{\alpha} ij}$ is the matrix element for the removal(addition) of an electron defined as
\begin{subequations}
\begin{alignat}{2}
    \Gamma^{Nr}_{{\alpha} ij}&=\sum_{\sigma\tau}\gamma_{\alpha}\left|\bra{N,i}\hat{c}_{{\alpha}\sigma\tau}\ket{N+1,j}\right|^2\\
    \Gamma^{Na}_{{\alpha} ij}&=\sum_{\sigma\tau}\gamma_{\alpha}\left|\bra{N,i}\hat{c}_{{\alpha}\sigma\tau}^\dagger\ket{N-1,j}\right|^2.
\end{alignat}
\end{subequations}
The coefficient $\gamma_{L(R)}$, illustrated in Fig.~\ref{fig:intro_fig}(b), represents the contact coupling rates of the DQD with the source(drain), given by  \cite{current_gamma}
\begin{align}
    \gamma_{\alpha} = 2\pi\sum_{k,\sigma,\tau} \left|\uptau_{\alpha k}\right|^2\delta(E-\epsilon_{\alpha k\sigma\tau}).\label{eq:gamma}
\end{align}
The above relation is obtained from the perturbative expansion of the density matrix equation using the spin and valley preserving contact tunneling Hamiltonian \cite{bhasky_main,bhasky_2,current_gamma,bhasky_3}
\begin{align}
    \hat{H}_{T} &= \sum_{\alpha,k,\sigma,\tau}\uptau_{\alpha k}\hat{d}^\dagger_{\alpha k\sigma\tau}\hat{c}_{\alpha\sigma\tau}+\text{h.c.},\label{eq:contact_tunneling_hamiltonian}
\end{align}%
where $\uptau_{\alpha k}$ with $\alpha=L(R)$ is the tunneling Hamiltonian matrix element between the $k^\text{th}$ eigenstate of the source(drain) and the left(right) QD, independent of the spin or the valley pseudo-spin. The tunneling between the contacts and the DQD preserves the spin $\sigma$ and the valley pseudo-spin $\tau$. The operator $d_{\alpha k\sigma\tau}\left(d^\dagger_{\alpha k\sigma\tau}\right)$ is the annihilation(creation) operator for an electron in the $k^\text{th}$ single electron eigenstate of the source ($\alpha=L$) or the drain ($\alpha=R$) with energy $\epsilon_{\alpha k\sigma\tau}$. 
To obtain the total current, one has to solve \eqref{eq:master_equation} for $\dot{P}^N_i=0\forall N,i$ under the constraint $\sum_{N,i}P^N_i=1$. The expression for current is given by \cite{bhasky_main}
\begin{align}
    I = \frac{e}{h}\sum_{N=1}^8\sum_{\langle i,j\rangle}\left[R^L_{(N-1,j)\rightarrow(N,i)}P^{N-1}_j-R^L_{(N,i)\rightarrow(N-1,j)}P^N_i\right].\label{eq:current}
\end{align}
\subsection{Current collapse mechanism\label{subsec:formalism:current_collapse}}
The mechanism of transport, discussed in Sec.~\ref{subsec:formalism:transport}, ensues two selection rules: (i) only states with $\Delta N=\pm1$ have non-zero transition rates between them, and (ii) the spin and the valley pseudo-spin must be preserved while tunneling. These two selection rules work together with Pauli's exclusion principle (no two electrons can have the same set of the three quantum numbers: dot, spin, and valley) to create the "dark" or blocking states, that in turn result in current blockades.

The $N=1$ and $N=2$ eigenspaces form an effective three-state model, as shown in Fig.~\ref{fig:ndr_physics}(a): four $B$ states in the $N=1$ space, and six $C$ states and six $D$ states in the $N=2$ space. If the magnetic field is not too high, then the spin-orbit coupling and the Zeeman splittings of the substates of the $C$ and the $D$ states are much smaller compared to the energy difference between the $C$ and the $D$ space.

We develop the notation $\mathcal{C}_{\zeta_2}\left(\mathcal{D}_{\zeta_2}\right)$ to denote a transition from the state $B_{\zeta_1}$ to the state $C_{\zeta_1\zeta_2}\left(D_{\zeta_1\zeta_2}\right)$ for all $\zeta_1\neq\zeta_2$, where $\zeta_1, \zeta_2$ are two spin-valley states. For a particular $\zeta_2$, there exist $3$ values of $\zeta_1$ corresponding to a transition from $B_{\zeta_1}$ to $C_{\zeta_1\zeta_2}$. It is trivial to check that the energy gap is the same for the $3$ transitions. Thus, each of the $B$ states can tunnel to $3$ possible $C$ states. Each of the $C$ states can tunnel to $2$ possible $B$ states. Likewise, each of the $B$ states can tunnel to $3$ possible $D$ states, and each of the $D$ states can tunnel to $2$ possible $B$ states. When all possible $\mathcal{C}$ and $\mathcal{D}$ transitions are accessible within the bias window, the master equation can be framed as
\begin{subequations}
\begin{alignat}{3}
\dot{P}_B &= -\left(R_{B\rightarrow C}+R_{B\rightarrow D}\right)P_B + R_{C\rightarrow B}P_C + R_{D\rightarrow B}P_D\\
\dot{P}_C &= -R_{C\rightarrow B}P_C + R_{B\rightarrow C}P_B\\
\dot{P}_D &= -R_{D\rightarrow B}P_D + R_{B\rightarrow D}P_B.
\end{alignat}
\end{subequations}
Previous works \cite{bhasky_main,dark_channels} have demonstrated that blockade occurs along the forward bias if
\begin{align}
    \frac{1}{R_{D\rightarrow B}}>\frac{1}{R_{B\rightarrow C}}+\frac{1}{R_{C\rightarrow B}},\label{eq:blockade_condition_forward}
\end{align}
and along the reverse bias if 
\begin{align}
    \frac{1}{R_{B\rightarrow D}}>\frac{1}{R_{B\rightarrow C}}+\frac{1}{R_{C\rightarrow B}}.\label{eq:blockade_condition_reverse}
\end{align}
The transport rates are given as a product of the transition matrix element and the coupling rates with the source or the drain. At very low temperatures, the Fermi-Dirac distribution assumes the form of the Heaviside theta function. In this regime, for forward bias,
\begin{subequations}
\begin{alignat}{4}
R_{B\rightarrow C}&=\gamma_LM^L_{BC}\\
R_{B\rightarrow D}&=\gamma_LM^L_{BD}\\
R_{C\rightarrow B}&=\gamma_RM^R_{CB}\\
R_{D\rightarrow B}&=\gamma_RM^R_{DB}.
\end{alignat}
\label{eq:rates}
\end{subequations}
For reverse bias, we simply swap the symbols $L$ and $R$. The matrix elements are evaluated using the electron creation and annihilation operators. In the case where the bias window encloses all the $\mathcal{C}$ and $\mathcal{D}$ transitions, these matrix elements take the form
\begin{subequations}
\begin{alignat}{8}
M^L_{BC} &= 3\left|\bra{C_{\zeta_1\zeta_2}}c^\dagger_{L\zeta_2}\ket{B_{\zeta_1}}\right|^2=3\left(\xi\beta+\eta\alpha\right)^2\\
M^L_{CB} &= 2\left|\bra{C_{\zeta_1\zeta_2}}c^\dagger_{L\zeta_2}\ket{B_{\zeta_1}}\right|^2=2\left(\xi\beta+\eta\alpha\right)^2\\
M^R_{BC} &= 3\left|\bra{C_{\zeta_1\zeta_2}}c^\dagger_{R\zeta_2}\ket{B_{\zeta_1}}\right|^2=3\left(\xi\alpha+\eta\kappa\right)^2\\
M^R_{CB} &= 2\left|\bra{C_{\zeta_1\zeta_2}}c^\dagger_{R\zeta_2}\ket{B_{\zeta_1}}\right|^2=2\left(\xi\alpha+\eta\kappa\right)^2\\
M^L_{BD} &= 3\left|\bra{D_{\zeta_1\zeta_2}}c^\dagger_{L\zeta_2}\ket{B_{\zeta_1}}\right|^2=3\eta^2\\
M^L_{DB} &= 2\left|\bra{D_{\zeta_1\zeta_2}}c^\dagger_{L\zeta_2}\ket{B_{\zeta_1}}\right|^2=2\eta^2\\
M^R_{BD} &= 3\left|\bra{D_{\zeta_1\zeta_2}}c^\dagger_{R\zeta_2}\ket{B_{\zeta_1}}\right|^2=3\xi^2\\
M^R_{DB} &= 2\left|\bra{D_{\zeta_1\zeta_2}}c^\dagger_{R\zeta_2}\ket{B_{\zeta_1}}\right|^2=2\xi^2.
\end{alignat}
\label{eq:matrix_elements}
\end{subequations}
Substituting the results from \eqref{eq:rates} and \eqref{eq:matrix_elements} into \eqref{eq:blockade_condition_forward} and \eqref{eq:blockade_condition_reverse}, and using the assumption $\gamma_L=\gamma_R$, we obtain the condition for realizing blockaded transport as
\begin{align}
    \begin{cases}
    \frac{1}{2\xi^2}>\frac{1}{3(\xi\beta+\eta\alpha)^2} + \frac{1}{2(\xi\alpha+\eta\kappa)^2} & \mu_L>\mu_R\\
    \frac{1}{3\eta^2}>\frac{1}{2(\xi\beta+\eta\alpha)^2} + \frac{1}{3(\xi\alpha+\eta\kappa)^2} & \mu_L<\mu_R.
    \end{cases}\label{eq:final_ndr_conditions}
\end{align}
Under the conditions stipulated in \eqref{eq:final_ndr_conditions}, a Pauli blockade is realized only when a $\mathcal{C}$ transition and its corresponding $\mathcal{D}$ transition counterpart are accessed within the bias window at the same time. At low magnetic fields, $C$ states typically have lower energy than $D$ states. Therefore, as the source-drain bias is gradually increased, $\mathcal{C}$ transitions are accessed first, without their $\mathcal{D}$ counterpart. With the gradual increase in the source-drain bias, the current initially rises due to the entry of a $\mathcal{C}$ state into the bias window and then falls with the entry of its $\mathcal{D}$ counterpart. For this reason, we have labeled the states $C$ ("conducting") and $D$ ("dark"). In principle, if there is a single transition accessible within the bias window, there will be an increase in current, irrespective of whether it is a $\mathcal{C}$ or a $\mathcal{D}$ transition. A blockade can therefore occur if and only if
\begin{enumerate}
    \item The conditions in \eqref{eq:final_ndr_conditions} are satisfied, AND
    \item A $\mathcal{C}$ and its corresponding $\mathcal{D}$ transition are both accessible in the bias window.
\end{enumerate}
When Zeeman splittings become comparable to the energy difference between the $C$ and the $D$ states, it becomes intractable to solve for the current analytically. In such cases, we resolve to numerical simulations, as we see in section~\ref{sec:results}. The optimal regime for qubit initialization and readout is the presence of Pauli blockade along one bias direction and its absence along the other along the other bias direction. Taking inspiration from literature \cite{bhasky_main}, we set $\varepsilon_L>\varepsilon_R$ and $\varepsilon_L+U_{ij}\approx\varepsilon_R+U_{ii}$, so that $\xi\ll\eta$ and $\beta\ll\alpha\approx\kappa$. Substitution of the aforementioned parameters into \eqref{eq:final_ndr_conditions} satisfies the blockade conditions along the forward bias ($\mu_L>\mu_R$), but not along the reverse bias ($\mu_L<\mu_R$). In performing our simulations, we shall, therefore, stick to this regime only. A concise description of the mathematical formulation for the current collapse mechanism in this regime is discussed in Appendix~\ref{sec:app2}.

There is a simple explanation for the $\mathcal{D}$ transitions causing blockade. The $C$ states are a superposition of $(1,1), (2,0), (2,0)$ occupancies while the $D$ states are just $(1,1)$. Owing to Pauli's exclusion principle, an electron in a $D$ state in a particular spin-valley configuration cannot tunnel into the $(0,2)$ (or $(2,0)$) component of the $C$ state with the same spin-valley configuration. We therefore encounter a current blockade from $(1,1)$ to $(0,2)$ (or $(2,0)$). As seen later in Fig.~\ref{fig:bias_triangles}, transitions from $(0,2)$ to $(1,1)$ show high current while the reverse shows little to no current.

`

\section{Results\label{sec:results}}
\begin{figure*}[thpb]
\centering
\captionsetup[subfigure]{oneside,margin={0.3cm,-1cm}}
\captionsetup[subfloat]{oneside,margin={0.3cm,-1cm},labelfont=bf}
\begin{minipage}{\linewidth}
    \centering
    \hspace{12mm}\includegraphics[width=0.40\linewidth]{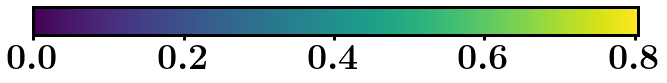}\hspace{20mm}
    \includegraphics[width=0.40\linewidth]{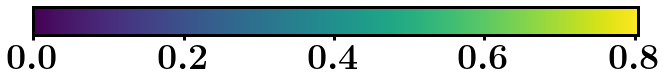}
\end{minipage}
\begin{minipage}{\linewidth}
    \centering
    \includegraphics[width=0.45\linewidth]{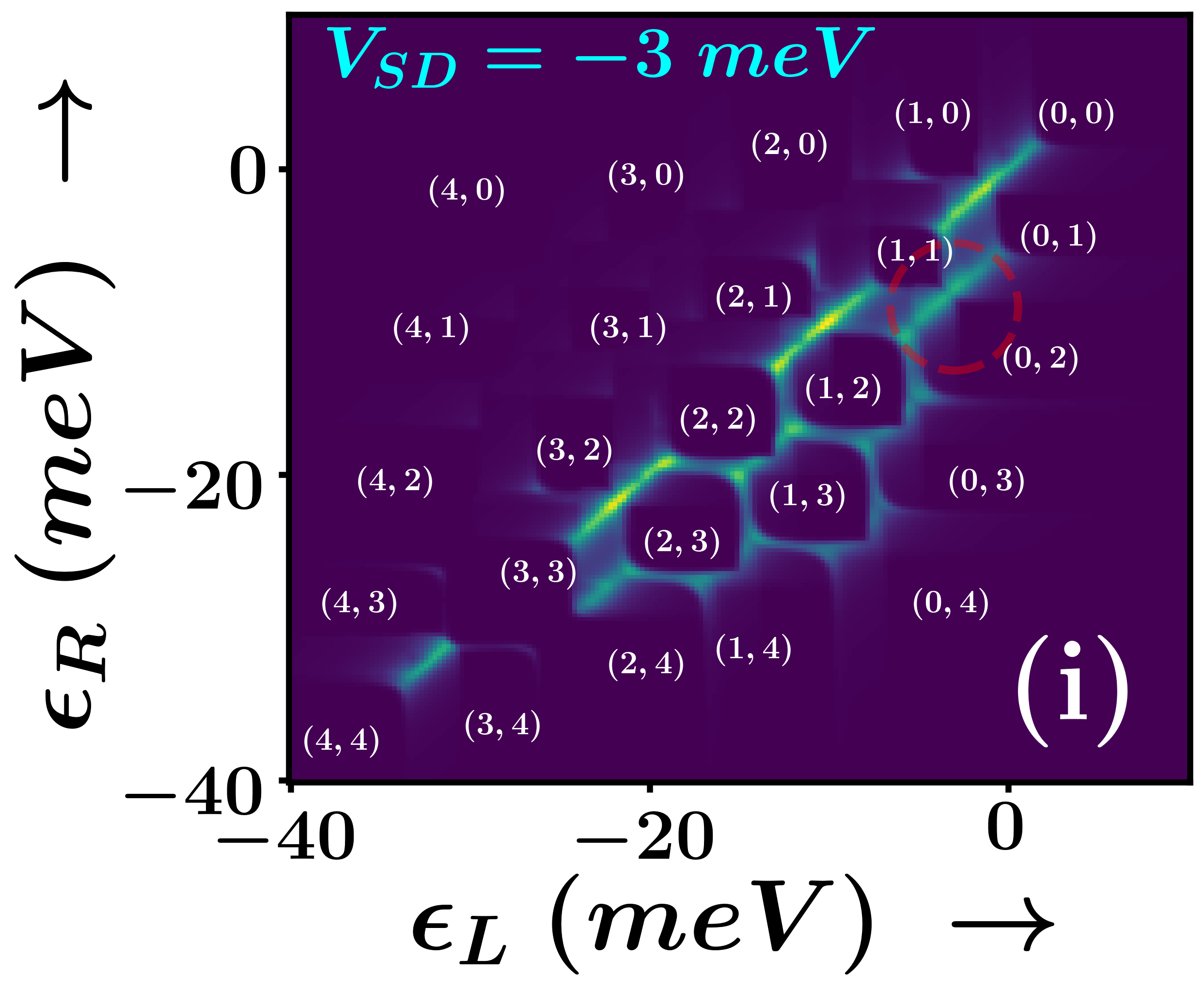}\hspace{10mm}
    \includegraphics[width=0.45\linewidth]{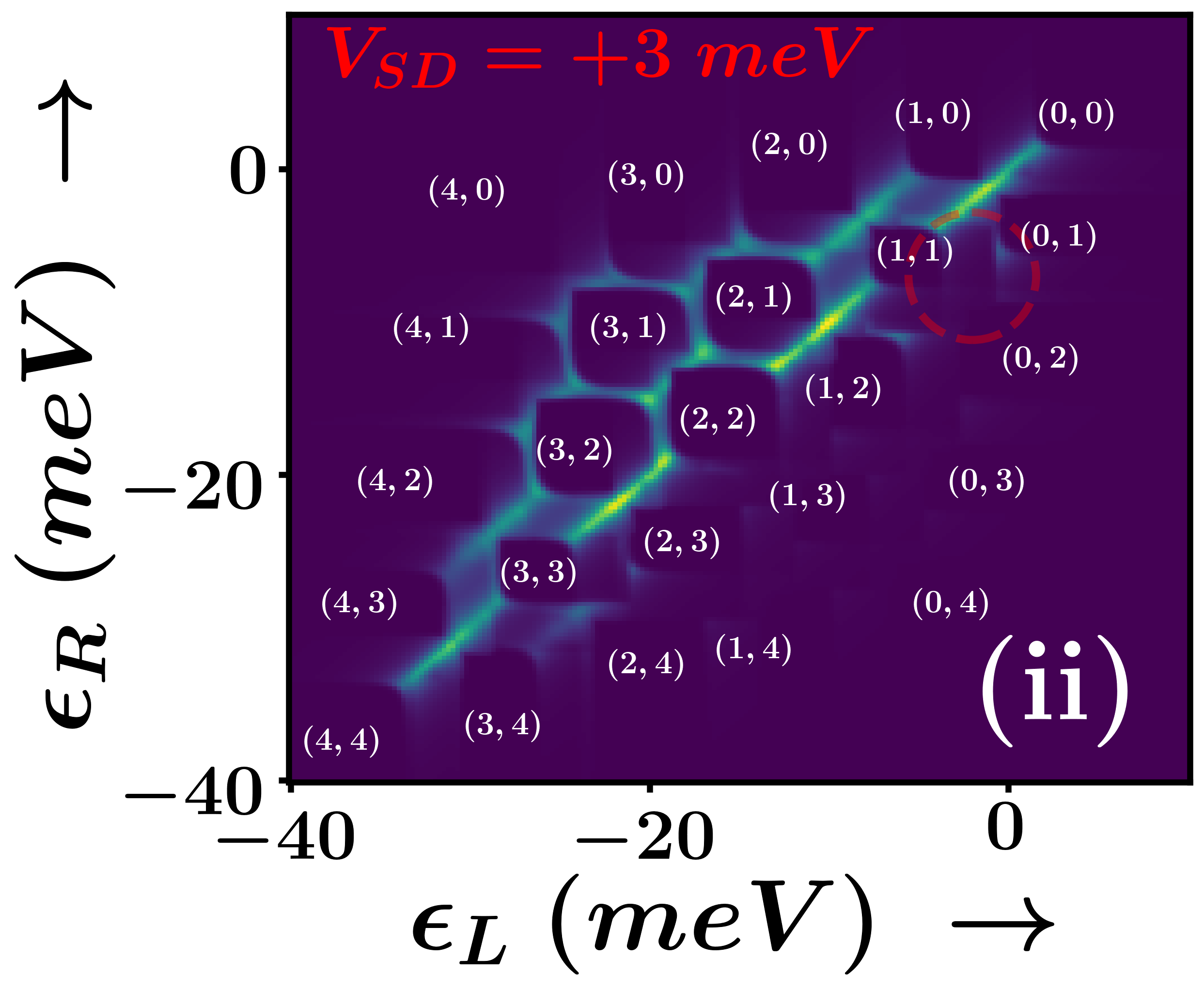}
\end{minipage}
\captionsetup{justification=raggedright,singlelinecheck=false}
\caption{
 Charge stability diagrams for (i) Reverse bias of $V_{SD}=-3$ meV and (ii) Forward bias of $V_{SD}=+3$ meV. The current measured is in nA. The external magnetic field is set to $1.0$ T. Notice that the occupancy is flipped as compared to expeimental sweep of the plungers: this is because plungers and onsite energies have negative correlations. The most probable charge occupancy number, in regions where there is negligible current, is indicated in white. Bias triangles occur at the inter-dot junction of $(1,1)-(0,2)$ as is indicated by the red circle.
}
\label{fig:charge_stability}
\end{figure*}

\begin{figure*}[thpb]
\centering
\captionsetup[subfigure]{oneside,margin={0.3cm,-1cm}}
\captionsetup[subfloat]{oneside,margin={0.3cm,-1cm},labelfont=bf}
%
\subfloat[]{
\begin{minipage}{0.33\linewidth}
    \hspace{10mm}\includegraphics[width=0.8\textwidth]{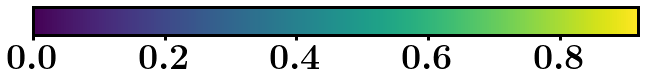}\\
    \includegraphics[width=\textwidth]{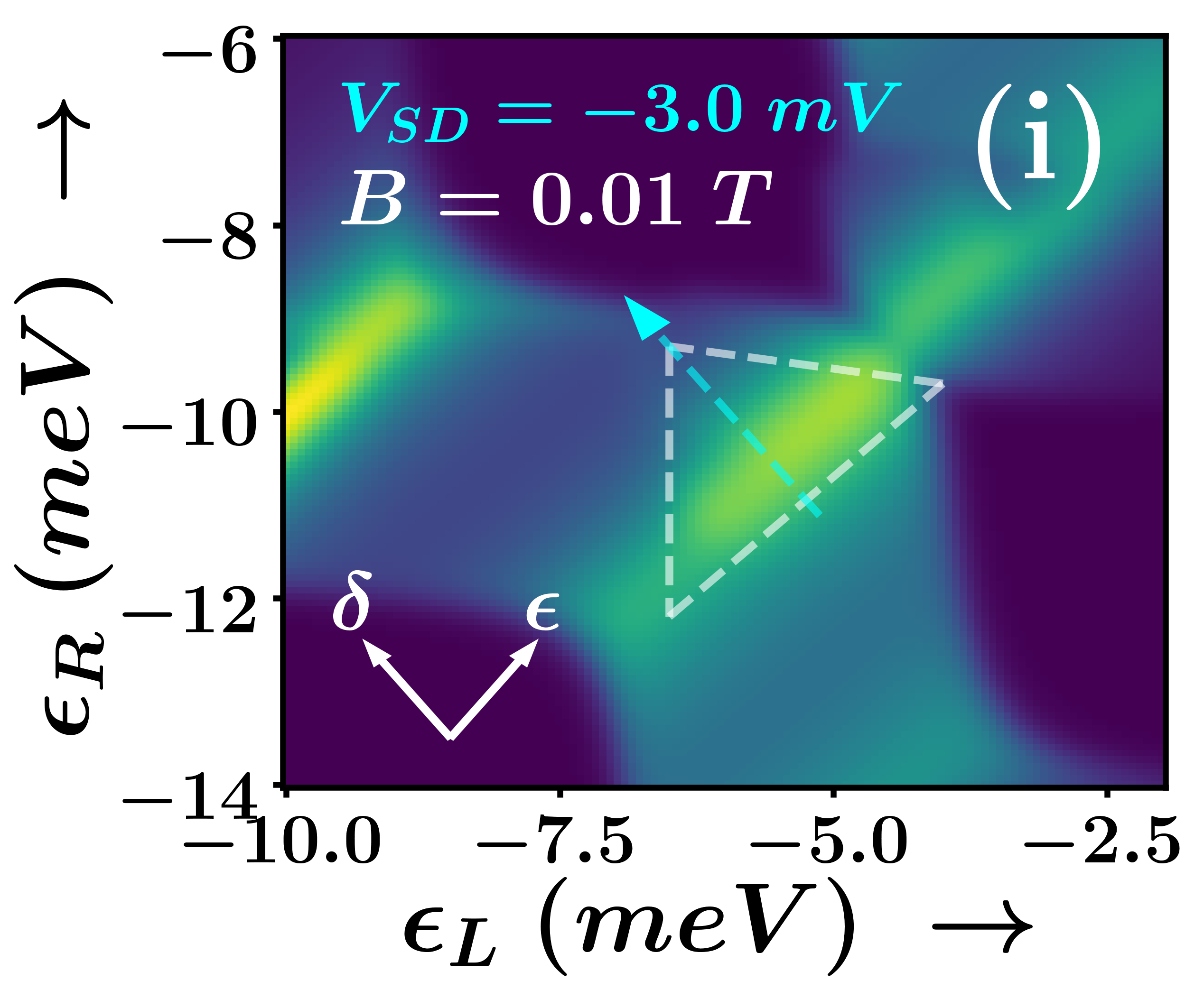}\\[4ex]
    \hspace{10mm}\includegraphics[width=0.8\textwidth]{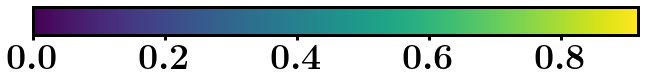}\\
    \includegraphics[width=\textwidth]{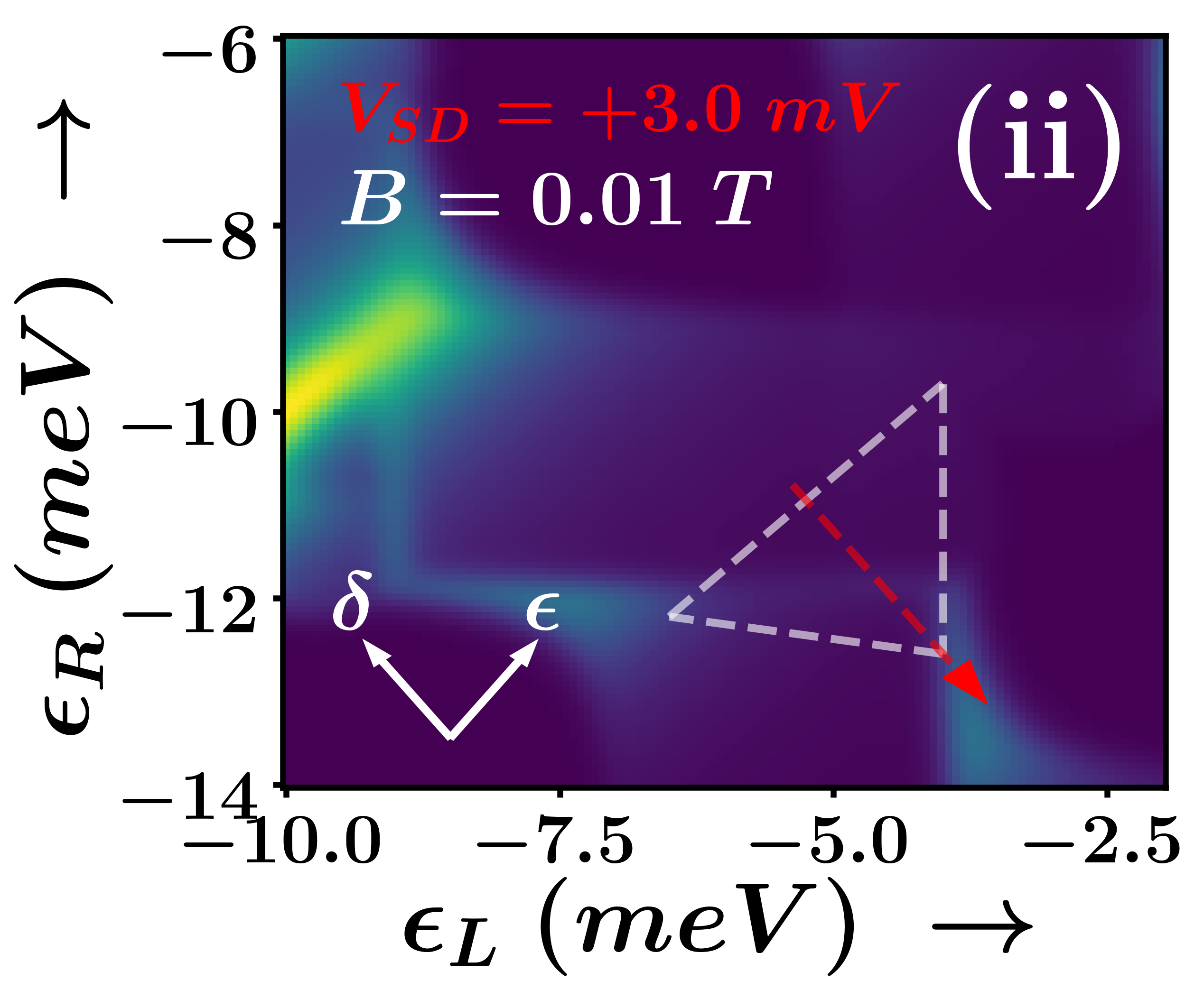}\\[4ex]
    \includegraphics[width=\textwidth]{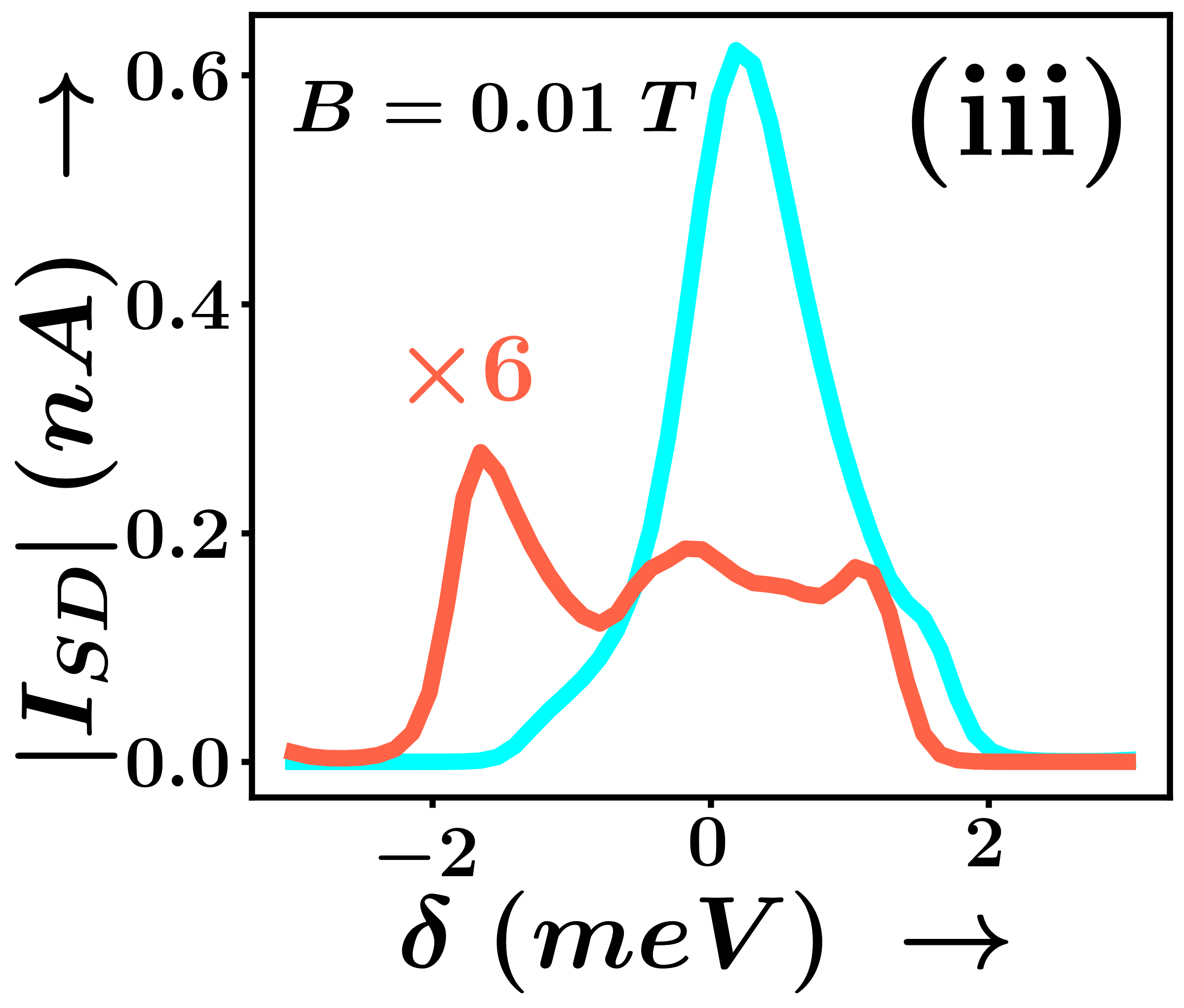}
\end{minipage}}
\subfloat[]{
\begin{minipage}{0.33\linewidth}
    \hspace{8mm}\includegraphics[width=0.8\textwidth]{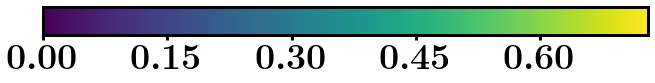}\\
    \includegraphics[width=\textwidth]{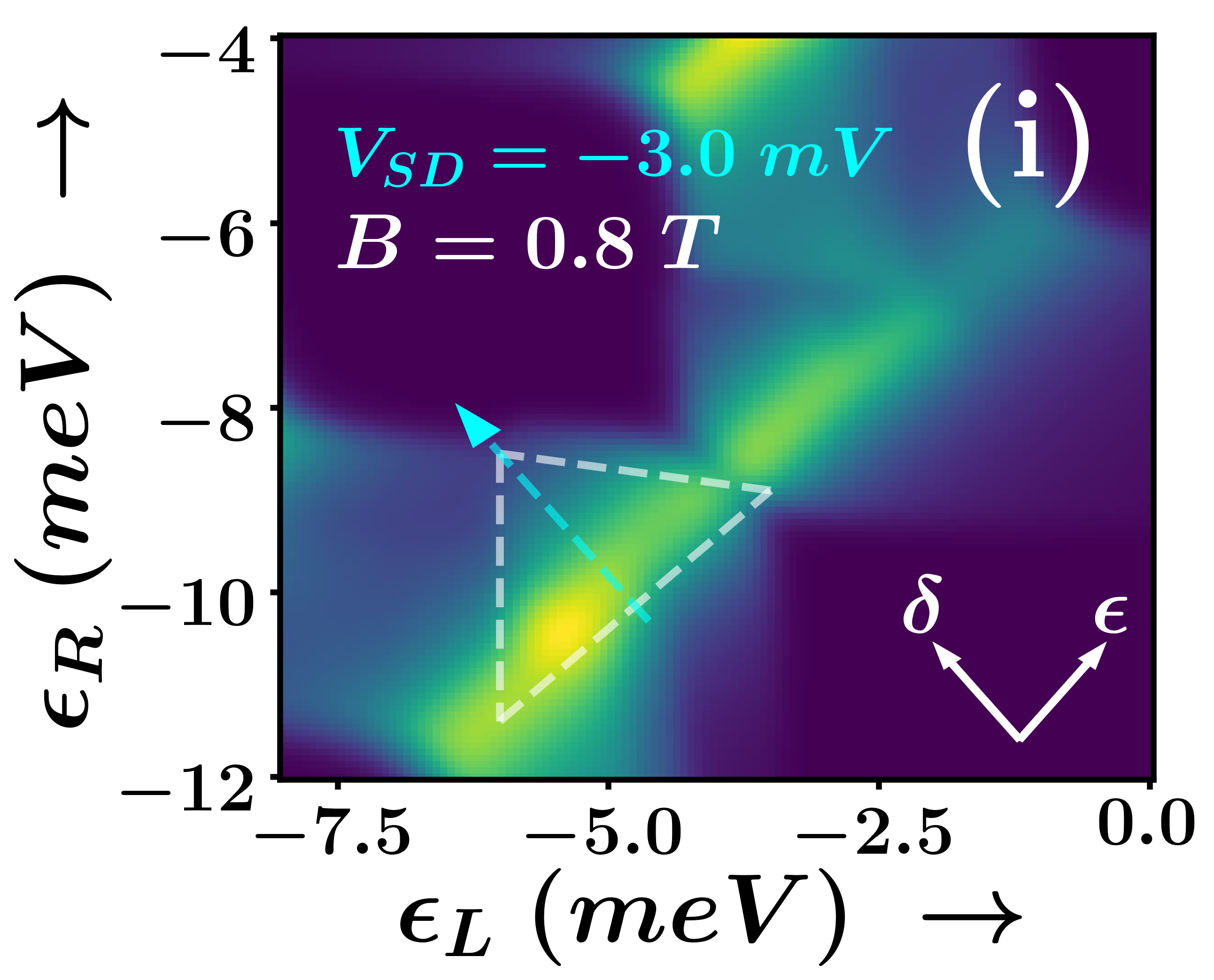}\\[4ex]
    \hspace{8mm}\includegraphics[width=0.8\textwidth]{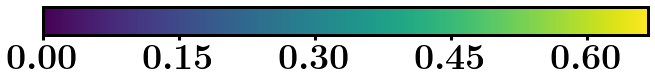}\\
    \includegraphics[width=\textwidth]{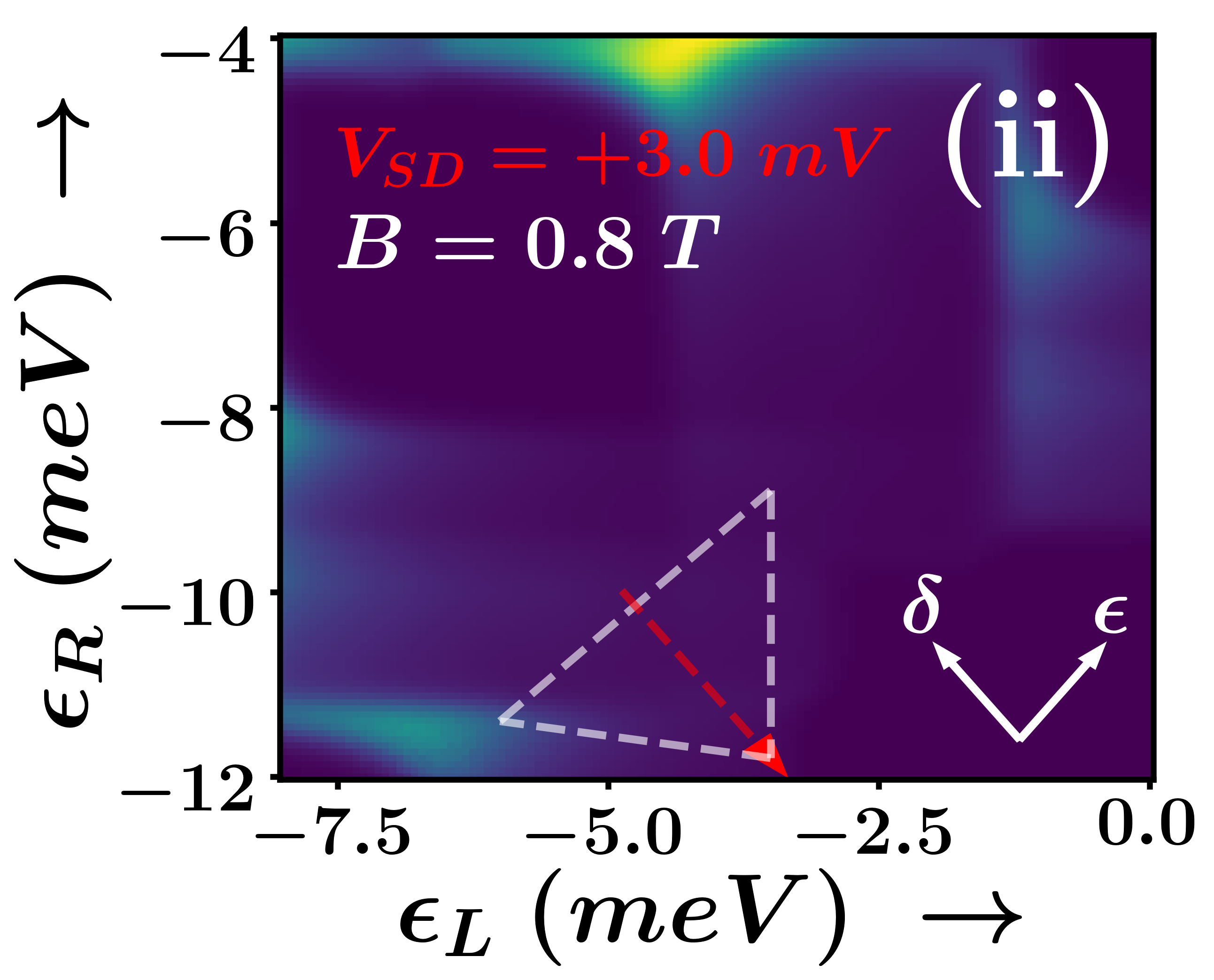}\\[4ex]
    \includegraphics[width=\textwidth]{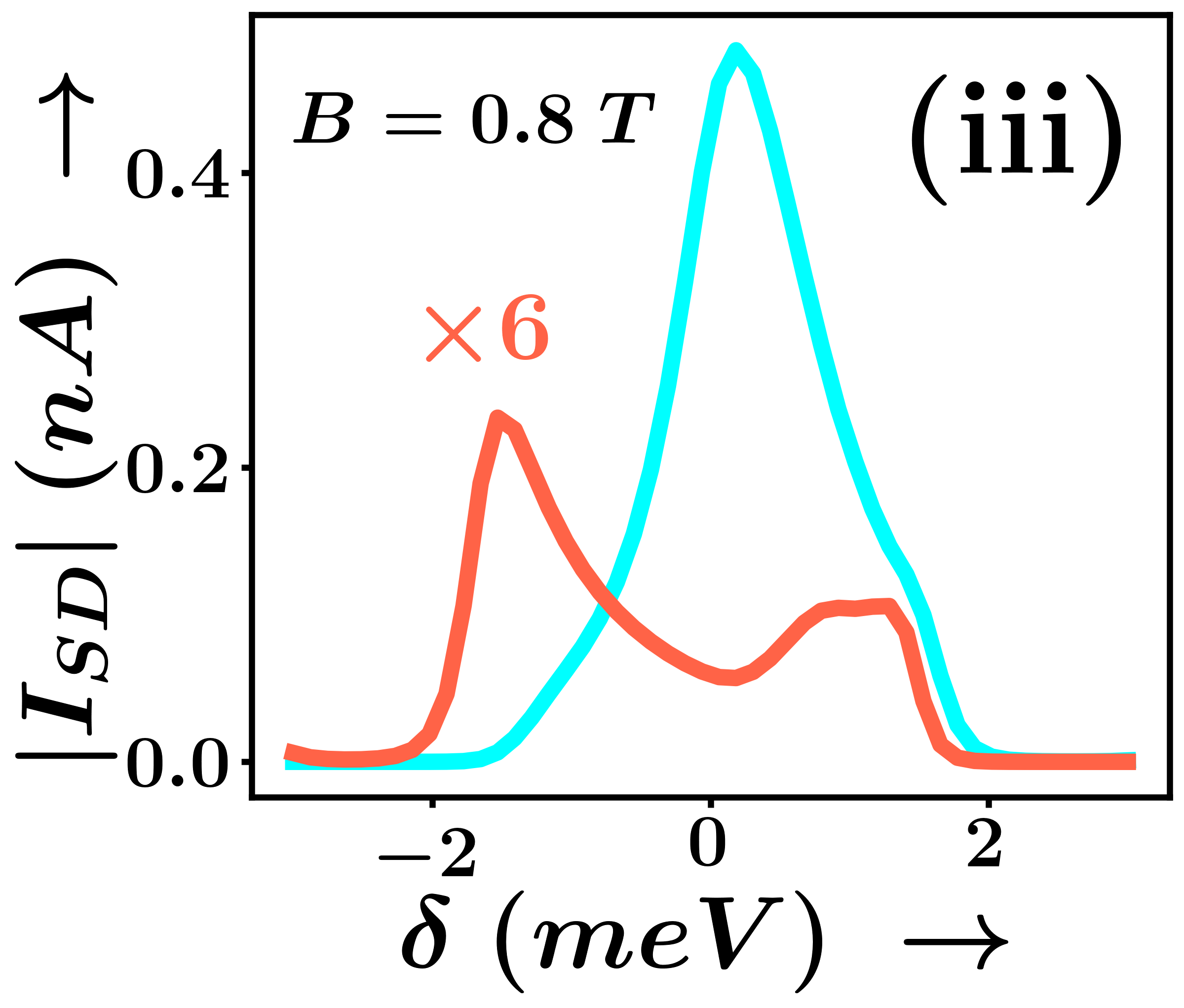}
\end{minipage}}
\subfloat[]{
\begin{minipage}{0.33\linewidth}
    \hspace{8mm}\includegraphics[width=0.8\textwidth]{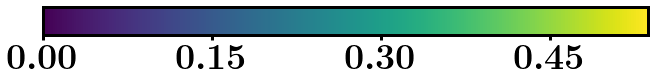}\\
    \includegraphics[width=\textwidth]{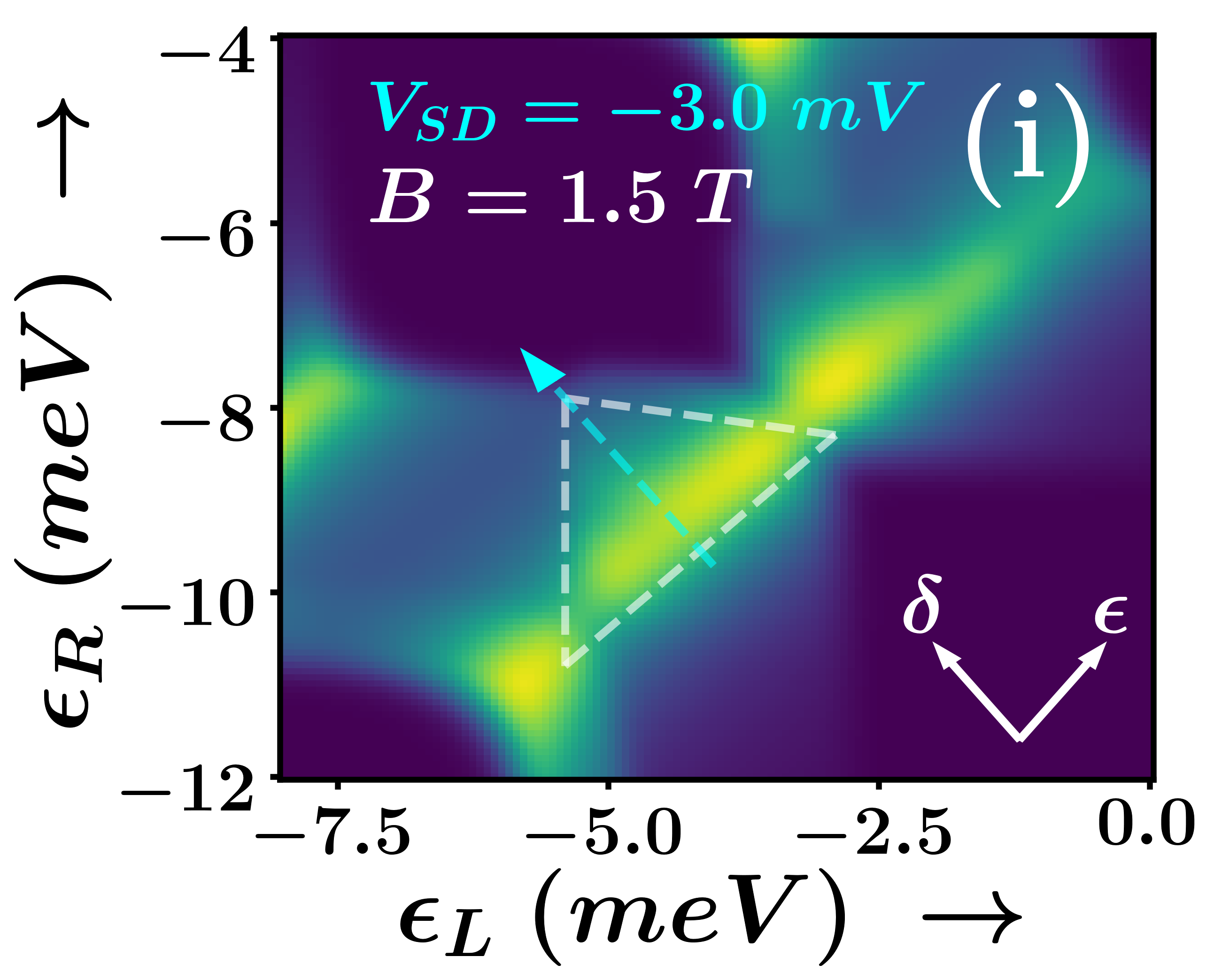}\\[4ex]
    \hspace{8mm}\includegraphics[width=0.8\textwidth]{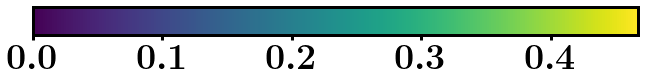}\\
    \includegraphics[width=\textwidth]{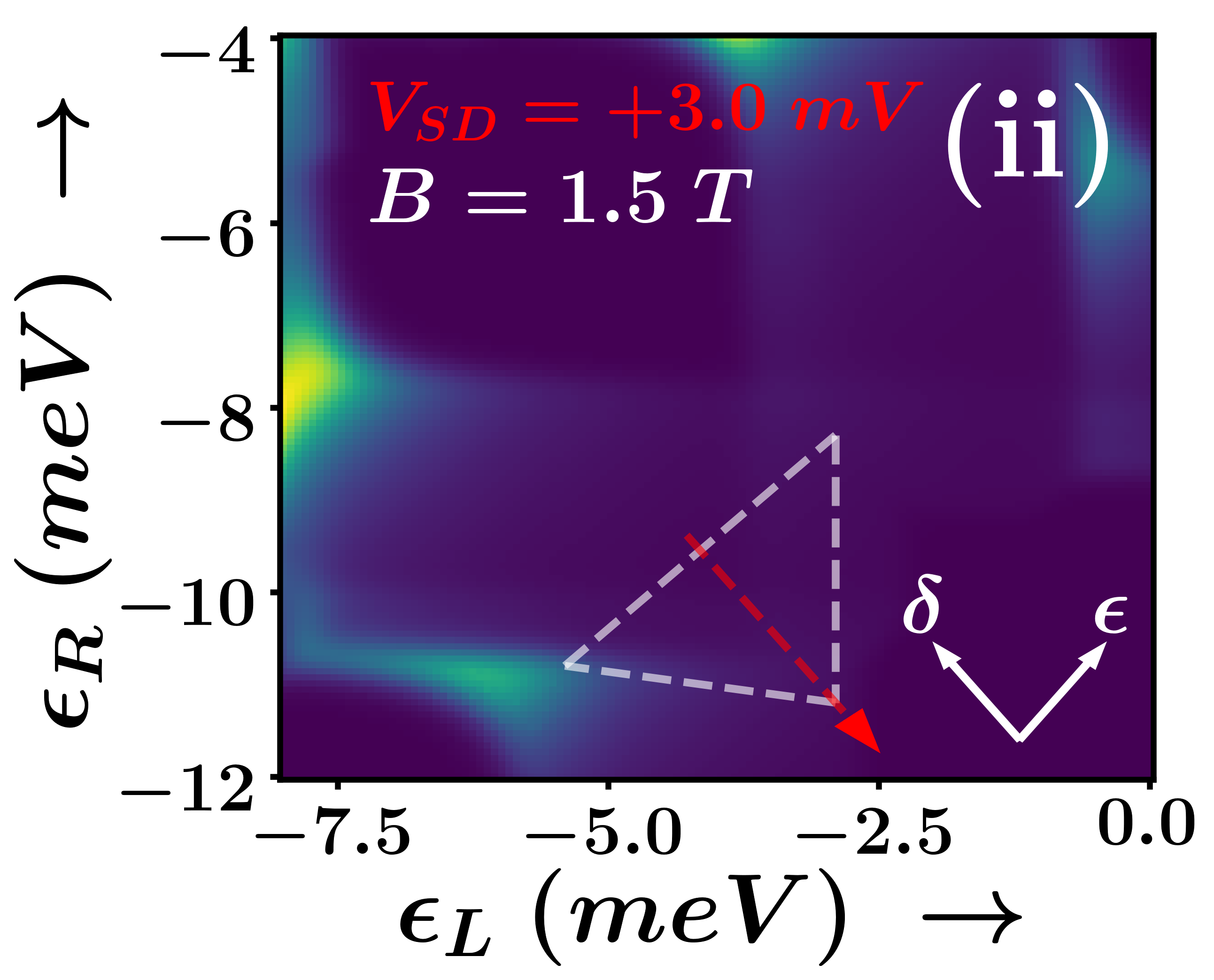}\\[4ex]
    \includegraphics[width=\textwidth]{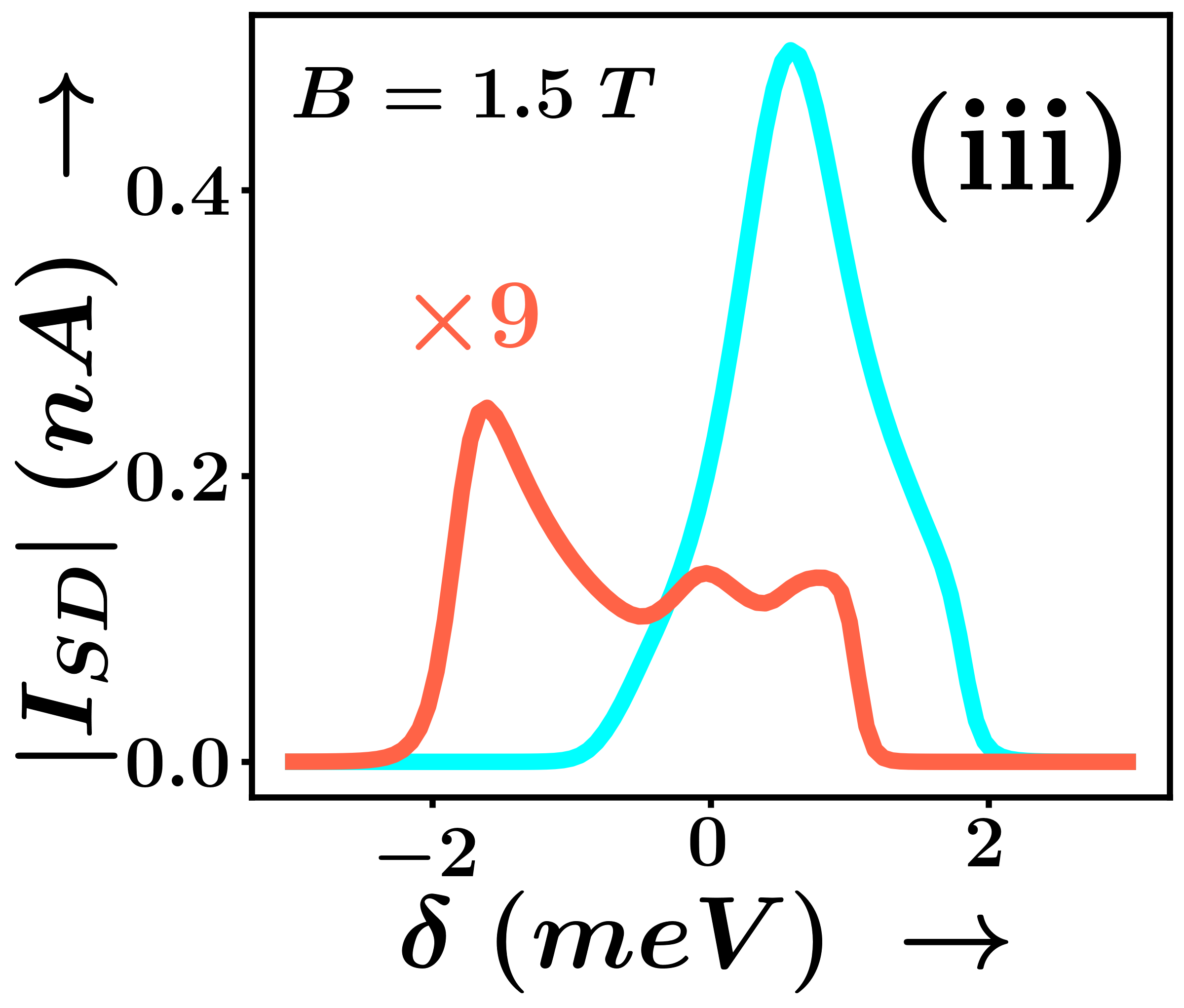}
\end{minipage}}
\captionsetup{justification=raggedright,singlelinecheck=false}
\caption{
(i) and (ii) in each figure show the bias triangles obtained as a result of sweeping the plunger-controlled onsite energies $\varepsilon_L$ and $\varepsilon_R$ for \textbf{(a)} $B=0.01$ T, \textbf{(b)} $B=0.8$ T, and \textbf{(c)} $B=1.5$ T. The current is measured in nA. Bias triangles obtained in the $(1,1)-(0,2)$ inter-dot transition region, as shown by the red circles in Fig.~\ref{fig:charge_stability} are identified by the dotted white lines. The blue(red) arrow shows the detuning axis for reverse(forward) bias, respectively. (iii) shows the current along the detuning axis $\delta$ for forward(red) and reverse(blue) biases of magnitude $3$ meV.
}
\label{fig:bias_triangles}
\end{figure*}

This section discusses how the parameters of the sample manifest themselves in determining the currents through the dots and creating regions of blockaded transport. We summarize the effects of (i) onsite energies, (ii) source-drain bias voltage, and (iii) magnetic field on the current flowing through the dots.

We use realistic values for the parameters of our model. The size of each lateral quantum dot typically lies in the range of $1-20$ nm \cite{DQD_size1,DQD_size2}. The dielectric constant of BLG is $\epsilon_r = 6\pm2$ \cite{DQD_permittivity}. Thus, the corresponding Coulomb repulsion energy lies in the range of $1-50$ meV \cite{TMDC_paper}. The gates have a thickness of the order of magnitude of the radii of the dots. The Coulomb repulsion is inversely proportional to the distance. Complying with these values, we set $U_{ii}=7$ meV and $U_{ij}=2$ meV. We shall discuss the Physics for weakly coupled quantum dots and hence, choose the inter-dot tunneling energy $t=-0.4$ meV.  The coupling rates are set as $\gamma_L=\gamma_R=5\times10^{-6}$. Experimentalists have measured $\Delta_{SO}\approx68-80$ meV \cite{main_exp_blg_2,main_exp_blg,so_exp_blg} and the g-factors to be $g_s\approx2$ and $g_v\approx15-30$ \cite{gf_1,gf_2,gf_3,gf_4,gf_5,gf_6,gf_7}. For our system, we consider $\Delta_{SO}=80$ meV, $g_s=2$, $g_v=15$. We also choose the other parameters in accordance with the discussion in Sec.~\ref{subsec:formalism:current_collapse}. With these values, we proceed to solve the Hamiltonian~\eqref{eq:fermi_hubbard_hamiltonian}. Unless otherwise mentioned, we set the temperature to $2.0$ K to observe the finite temperature effects, such as current leakage.

\subsection{Current blockade\label{subsec:results:blockade}}
The mechanism for current blockade has been discussed extensively in section~\ref{subsec:formalism:current_collapse}. Current blockade arises when (i) transitions from the bonding state to both the conducting and the corresponding dark states are available within the bias window, and (ii) The parameters of the system are such that the pair of these two available transitions is blocking. 

The magnetic field is first set at $B=0.01$ T so that the energy difference between the $C$ state and its corresponding $D$ state is significantly higher than that between the substates of $C$ or $D$ states. The temperature is set as low as $30$ mK to observe flat current plateaus and sharp transitions between them. Fig.~\ref{fig:ndr_physics}(a) shows the conducting and blocking transitions. As the orange $\mathcal{C}$ transitions to the orange $C$ state is accessed, the current rises. Opening the bias window a little more allows transition into both the orange $C$ and $D$ states, in which case, we observe a current blockade. As the bias is increased even more, we observe the entry of the green $C$ state transition into the bias window, followed by the green $D$ state transition, thereby first increasing and then subsequently decreasing the total current. The corresponding effect on the current is summarized in Fig.~\ref{fig:ndr_physics}(b).

Correlating the transitions and the source-drain bias levels depicted in Fig.~\ref{fig:ndr_physics}(c) with the current shown in Fig.~\ref{fig:ndr_physics}(b)(i), we can explain the causes and regimes of Pauli blockades. In the forward bias, whenever a new $\mathcal{C}$ transition enters the bias window, but its corresponding $\mathcal{D}$ transition does not, we observe a rise in current. When the corresponding $\mathcal{D}$ transition enters the window, we get a sudden drop in current, indicating that the $D$ state causes a blockade. The pink line cut at $V_{SD}\approx0$ encloses no transition within its bias window, and thus the current is $0$. At the brown line cut, two $\mathcal{C}$ transitions, $\mathcal{C}_{\kup}$ and $\mathcal{C}_{\kpdown}$, enter the bias window and therefore a sharp rise in current is observed. At the dark green cut (Fig.~\ref{fig:ndr_physics}(b)(ii)), we observe another rise in current due to the entry of a new $\mathcal{C}$ transition ($\mathcal{C}_{\kdown}$) into the bias window, without the entry of the corresponding $\mathcal{D}$ transition ($\mathcal{D}_{\kdown}$). At the ochre line cut, the first $\mathcal{D}$ transition ($\mathcal{D}_{\kpup}$) enters the bias window. Since $\mathcal{C}_{\kpup}$ is also present in the window, we observe a sharp dip in current. As $V_{SD}$ is increased further, more blocking transitions $\mathcal{D}$ enter the bias window, further reducing the current, as is shown by the purple line cut in Fig.~\ref{fig:ndr_physics}(b)(iii). When all the $\mathcal{C}$ and their corresponding $\mathcal{D}$ transitions are within the bias window, the current becomes negligible (gray cut). In the reverse bias, however, there is no blocking, and thus, both $\mathcal{C}$ and $\mathcal{D}$ transitions contribute positively to current.

\subsection{Charge stability diagram and bias triangles\label{subsec:results:charge_stability_bias_triangles}}
Fig.~\ref{fig:charge_stability} shows the charge stability diagram, with regions of interest enclosed by a dotted red circle. Regions of high current represent a change in dot occupancy. Low current regions are formed when there is a large energy difference between the ground state and the excited states, and the most probable occupancy of each dot, $(n_L,n_R)$, in such regions is indicated within the figure. The occupancy of each dot is dictated by the most probable ground state. The charge stability plot serves to identify the region of the bias triangles. We focus our study on the region of $(1,1)-(0,2)$ inter-dot transition. Fig.~\ref{fig:bias_triangles} shows the bias triangles in this region. The bias triangles are spanned by two parameters: $\epsilon$ and $\delta$, as shown in Fig.~\ref{fig:bias_triangles}. $\epsilon=\varepsilon_L + \varepsilon_R$, upto a constant as $\varepsilon_L$ and $\varepsilon_R$ vary. Likewise, $\delta=\varepsilon_L - \varepsilon_R$, up to a constant. As can be seen from Fig.~\ref{fig:charge_stability}, $\epsilon$ controls the total onsite energy of the two dots together and hence fixes $N$, the total occupancy of the Fock subspace. We adjust $\epsilon$ so that we are in the $N=2$ subspace. The constant for $\delta$ is set to $\delta=0$ at the point where the current is maximum. Changing $\delta$ changes the transition energies of $\mathcal{C}$ and $\mathcal{D}$, thereby changing the current. 

\begin{figure}[t!]
\begin{minipage}{\linewidth}
\centering
\captionsetup[subfigure]{oneside,margin={0.3cm,-1cm}}
%
%
\begin{minipage}{\linewidth}
    \hspace{-12mm}\includegraphics[width=0.9\textwidth]{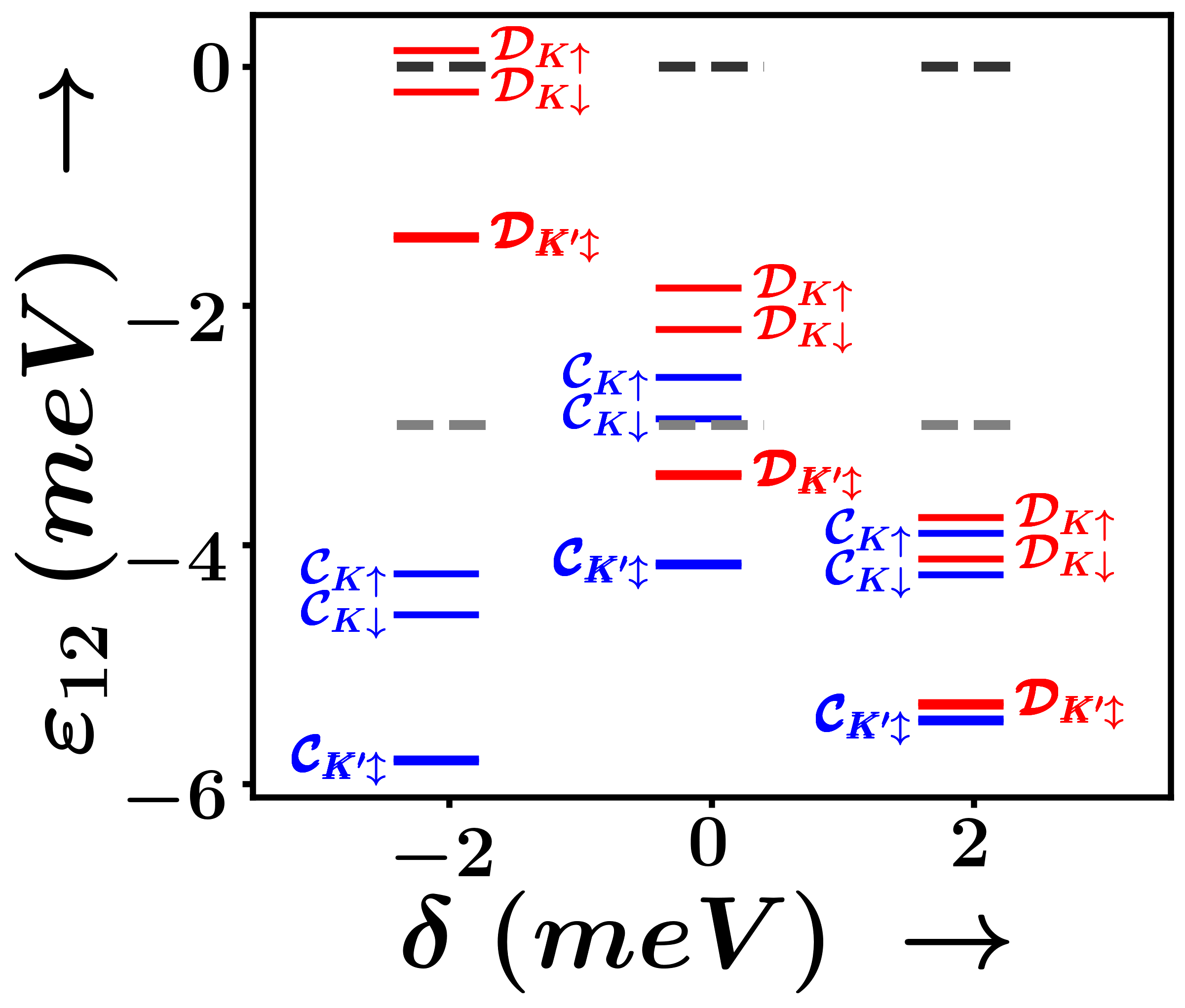}
\end{minipage}
\captionsetup{justification=raggedright,singlelinecheck=false}
\caption{
Energy transitions corresponding to the case presented in Fig.~\ref{fig:bias_triangles}(b) for $B=0.8$ T. 
The blue lines indicate the conducting states, and the red lines indicate the dark states. The gray lines indicate the source and drain chemical potentials, source being higher for forward bias and lower for reverse bias.
}
\label{fig:bias_traingle_states}
\end{minipage}
\end{figure}

We perform simulations for $B=0.01,0.8,1.5$ T in Fig~\ref{fig:bias_triangles}. An increase in the magnetic field causes a decrease in current along both bias directions. At large magnetic fields, the spin and valley Zeeman splittings are significant, therefore making it difficult to enclose many transitions within the bias window. This leads to an overall decrease in the total current across the DQD. To study the effect of detuning on the current, consider the case $B=0.8$ T. In the forward bias, at $\delta=-2$ meV, the $\mathcal{D}$ transitions are accessed within the bias window but not the $\mathcal{C}$ transitions. Thus, there is no current blockade (blockade occurs only when both $\mathcal{C}$ and $\mathcal{D}$ transitions are present), and we observe increasing current. At $\delta=0$, we have the maximum possible number of coupled $\mathcal{C}$ and $\mathcal{D}$ transitions (corresponding to the $\kpup$ and $\kdown$ transitions), thereby leading to a current dip (the current is still finite since blockade implies a sharp decrease in current and not necessarily $0$ current). Finally, as $\delta$ is increased, transitions accessible within the bias window decouple, leading to increasing current, till all available transitions leave the bias window at around $\delta=+2$ meV when the current drops to $0$. In the reverse bias case the current is simply peaked when the number of transitions accessible within the bias window is the maximum, which happens at $\delta=0$. As $\delta$ changes on either side of $0$, current falls as states keep exiting the bias window. As the value of $\delta$ is increased from $0$, the $\mathcal{D}$ transitions decrease in energy, while the $\mathcal{C}$ transitions increase in energy. Fig.~\ref{fig:bias_traingle_states} shows this effect in action. There is a small but finite current in the system even when there are no accessible transitions in the bias window, owing to leakage at finite temperature of the system.

\section{Conclusion\label{sec:conclusion}}
In this paper, we constructed a model capturing the delicate interplay of Coulomb interactions, inter-dot tunneling, Zeeman splittings, and intrinsic spin-orbit coupling in a DQD to simulate the Pauli blockades. Analyzing the relevant Fock-subspaces of the generalized Hamiltonian, coupled with the density matrix master equation technique for transport across the setup, we identified the generic class of blockade mechanisms. Most importantly, and contrary to what is widely recognized, we have shown that conducting and blocking states responsible for the Pauli-blockades are a result of the coupled effect of all degrees of freedom and cannot be explained using the spin or the valley pseudo-spin alone. We then numerically predicted the regimes where Pauli blockades might occur, and, to this end, we verified our model against actual experimental data and proposed that our model can be used to generate data sets for different values of parameters with the ultimate goal of training on a machine learning algorithm. Our work thus provides an enabling platform for a predictable theory-aided experimental realization of single-shot readout of the spin and valley states on DQDs based on 2D-material platforms.
\section*{Acknowledgements}
The author BM acknowledges the Visvesvaraya Ph.D Scheme of the Ministry of Electronics and Information Technology (MEITY), Government of India, implemented by Digital India Corporation (formerly Media Lab Asia). The author BM also acknowledges the support by the Science and Engineering Research Board (SERB), Government of India, Grant No. STR/2019/000030, and the Ministry of Human Resource Development (MHRD), Government of India, Grant No. STARS/APR2019/NS/226/FS under the STARS scheme.


\appendix
\section{Characterization of the Fock space\label{sec:app1}}

Consider the Hamiltonian defined in \eqref{eq:fermi_hubbard_hamiltonian}. We solve for the eigenstates in the $N=1$ and $N=2$ Fock subspaces.

\subsection{$\mathbf{N=1}$ Fock space}
The Hamiltonian takes the form of an $8\times8$ matrix. To diagonalize this Hamiltonian, start with the ansatz that the eigenstates are given by
\begin{align}
    \ket{\psi}=\xi\ket{L_\zeta}+\eta\ket{R_\zeta}
\end{align}
where $\zeta\in\{\kup, \kdown, \kpup, \kpdown\}$. 
Applying the Hamiltonian to $\ket{\psi}$ and solving for the eigenstates, we obtain the effective matrix to be diagonalized as
\begin{align}
M^\prime_1=
    \begin{pmatrix}
    \Tilde{\varepsilon}_L & t\\
    t & \Tilde{\varepsilon}_R
    \end{pmatrix}\label{eq:app:N=1}
\end{align}
where $\Tilde{\varepsilon}_L$ and $\Tilde{\varepsilon}_R$ are the effective onsite energies given as $\Tilde{\varepsilon}_\alpha=\varepsilon_\alpha+h_S+h_V+\Delta_{SO}$ where $h_S,h_V,\Delta_{SO}$ depend on the state $\zeta$ on the dot $\alpha\in\{L,R\}$. Replacing $\Tilde{\varepsilon}_\alpha$ with $\varepsilon_\alpha$ in \eqref{eq:app:N=1} will not change the eigenstates as all the extra terms contribute only along the diagonal. Since we are currently interested only in the eigenstates, we can make such a transformation. Thus, all four possibilities of $\zeta$ result in the same values of ($\xi$, $\eta$). It is trivial to check that that if a pair ($\xi$, $\eta$) forms an eigenstate of \eqref{eq:app:N=1}, so does ($\xi$, $-\eta$) (where $\xi,\eta\geq0$). Since $t<0$, the energy of ($\xi$, $\eta$) is lower and thus the state is called bonding, while the energy of ($\xi$, $-\eta$) is higher and the corresponding state is called anti-bonding state.

\subsection{$\mathbf{N=2}$ Fock space}
\begin{table}[ht]
\centering
\begin{tabular}{|l|l|l|l|l|} 
\hline
\textbf{State} & \textbf{Spin} & \textbf{Valley} & \textbf{SO} & \textbf{\#States} \\ 
\hline
$C_{\kup\kdown}$ & $0$ & $+1$ & $0$ & 3 \\
$C_{\kup\kpup}$ & $+1$ & $0$ & $0$ & 3 \\
$C_{\kup\kpdown}$ & $0$ & $0$ & $+2\Delta$ & 3 \\
$C_{\kdown\kpup}$ & $0$ & $0$ & $-2\Delta$ & 3 \\
$C_{\kdown\kpdown}$ & $-1$ & $0$ & $0$ & 3 \\
$C_{\kpup\kpdown}$ & $0$ & $-1$ & 0 & 3 \\ 
\hline
$D_{\kup\kdown}$ & $0$ & $+1$ & $0$ & 1 \\
$D_{\kup\kpup}$ & $+1$ & $0$ & $0$ & 1 \\
$D_{\kup\kpdown}$ & $0$ & $0$ & $+2\Delta$ & 1 \\
$D_{\kdown\kpup}$ & $0$ & $0$ & $-2\Delta$ & 1 \\
$D_{\kdown\kpdown}$ & $-1$ & $0$ & $0$ & 1 \\
$D_{\kpup\kpdown}$ & $0$ & $-1$ & 0 & 1 \\ 
\hline
$P_{\kup}$ & $+1$ & $+1$ & $+2\Delta$ & 1 \\
$P_{\kdown}$ & $-1$ & $+1$ & $-2\Delta$ & 1 \\
$P_{\kpup}$ & $+1$ & $-1$ & $-2\Delta$ & 1 \\
$P_{\kpdown}$ & $-1$ & $-1$ & $+2\Delta$ & 1 \\
\hline
\end{tabular}
\caption{Table for the states of $N=2$ Fock space}
\label{tab:N=2}
\end{table}

The $N=2$ space is described by a $28\times28$ Hamiltonian matrix. Diagonalization of such a matrix is hard. However, we know some ansatz that might help. The states have been classified according to the ansatz used in section~\ref{subsec:formalism:fock}. 

Let us start with the $C$ states. We have six $C$ states corresponding to the $\zeta_1\zeta_2$ combinations in $\{\kup\kdown,\ \kup\kpup,\ \kup\kpdown,\ \kdown\kpup,\ \kdown\kpdown,\ \kpup\kpdown\}$. These states are defined by
\begin{align}
\ket{C_{\zeta_1\zeta_2}} &= \alpha\left(\ket{L_{\zeta_1}R_{\zeta_2}}-\ket{L_{\zeta_2}R_{\zeta_1}}\right)\nonumber\\
&\qquad+\beta\ket{L_{\zeta_1}L_{\zeta_2}}+\kappa\ket{R_{\zeta_1}R_{\zeta_2}}
\end{align}
We use a similar approach as we did in the $N=1$ case. We set the onsite energies to $\Tilde{\varepsilon}_L$ and $\Tilde{\varepsilon}_R$ given as $\Tilde{\varepsilon}_\alpha=\varepsilon_\alpha+h_S+h_V+\Delta_{SO}$ where $h_S,h_V,\Delta_{SO}$ depend on the states $\zeta_1$ and $\zeta_2$ and $\alpha\in\{L,R\}$. For example, $\zeta_1\zeta_2=\kup\kdown$ has $h_S=0,h_V=+\mu_Bg_vB,\Delta_{SO}=0$ (calculated by summing the values of spin-Zeeman, valley-Zeeman and spin-orbit splittings respectively). Again, the eigenstates are invariant under such transformations and we end up with an effective matrix
\begin{align}
M^\prime_{2C}=
    \begin{pmatrix}
    \varepsilon_L + \varepsilon_R + U_{ij}  & t & t\\
    2t & 2\varepsilon_L+U_{ii} & 0\\
    2t & 0 & 2\varepsilon_R+U_{ii}
    \end{pmatrix}\label{eq:app:N=2}
\end{align}
corresponding to the eigenvector $(\alpha, \beta, \kappa)$ that we need to diagonalize. Note that this matrix has three non-degenerate eigenstates for each of the six $\zeta_1\zeta_2$ combinations. This is what gives rise to the total of eighteen $C$ states as we obtained in section~\ref{subsec:formalism:fock}.

We now move on to the $D$ states, which take the form $\ket{D_{\zeta_1\zeta_2}} = \frac{1}{\sqrt{2}}\left(\ket{L_{\zeta_1}R_{\zeta_2}}+\ket{L_{\zeta_2}R_{\zeta_1}}\right)$, as in \eqref{eq:N_2_fockspace}. It is easy to see that these are indeed the eigenstates of the Hamiltonian. Since there are six possible $\zeta_1\zeta_2$ combinations, we get six $D$ states. Finally, the $P$ states in \eqref{eq:N_2_fockspace} are the easiest to show as eigenstates of Hamiltonian.

Next, we shall see how these $C$ and $D$ states correlate with the standard notation of singlet and triplet. Table~\ref{tab:N=2} summarizes the states in the $N=2$ Fock space. The column Spin denotes the total spin of the state, while the column valley denotes the total valley pseudo-spin of the state. The SO column mentions the total spin-orbit splitting the state faces and \#States denotes the number of states corresponding to that particular label. The total spin is computed by adding the spins present in the system: same goes for the valley pseudo-spin. A state is called a spin singlet if its total spin is $0$ and the corresponding state is antisymmetric. It is called a triplet if it has total spin $\pm1$ or is symmetric with total spin $0$. Let us label singlets by $S$ and triplets by $T$. The triplets are labelled $T^0,T^{+},T^{-}$ depending on the total spin. The valley singlet and triplet are defined accordingly by looking at the pseudo-spin. Thus, the state $C_{\kup\kpup}$ is a spin-triplet-valley-singlet ($T_s^+S_v$) state, while $C_{\kpup\kpdown}$ is a spin-singlet-valley-triplet ($S_sT_v^-$) state. Here, we conclude that $C$ states are of the form $S_sT_v^+, S_sT_v^-, T_s^+S_v, T_s^-S_v, S_sS_v$. Notice that we have 2 $S_sS_v$ states, that are distinguished by their spin-orbit coupling. Likewise, $D$ states have the forms $T_s^0T_v^+, T_s^0T_v^-, T_s^+T_v^0, T_s^-T_v^0, T_s^0T_v^0$, with 2 $T_s^0T_v^0$ states, distinguished by SO coupling. $P$ states are of the forms $T_s^+T_v^+, T_s^+T_v^-, T_s^-T_v^+, T_s^-T_v^-$. The singlet states stay almost constant with respect to the magnetic field applied, while the triplet states change with $B$ linearly with a slope proportional to the spin (or valley pseudo-spin) and the corresponding g-factor. Fig.~\ref{fig:intro_fig}(d) summarizes these effects.  Note that such singlet-triplet notations, however, (i) are misleading about the occupancy of each dot and (ii) do not shed any light on the dynamics of current. We shall therefore refrain from using such notations and stick to the $C$, $D$, and $P$ formalism that we have developed.

\section{Current Blockade Calculations\label{sec:app2}}
We present a somewhat detailed calculation of the value of the current. To ease our calculations, we shall consider the $B$ states to be degenerate, and occurring with equal probability since the energy differences between them are negligible compared to those between the sub-states in the $C$ and $D$ states. We use $P_B$ to indicate the probability of each $B$ state. Let us denote the $i^\text{th}$ state in $B$, $C$, and $D$ as $\ket{B,i}$, $\ket{C,i}$, and $\ket{D,i}$ respectively. Let us denote the rate for an allowed transition from $\ket{B,i}$ to $\ket{C,j}$ as $R^\alpha_{BC}$, where $\alpha\in\{L,R\}$ indicates whether the source ($\alpha=L$) or the drain ($\alpha=R$) contributes to the transition. Likewise, we use $R^\alpha_{CB}$, $R^\alpha_{BD}$, and $R^\alpha_{DB}$ for transitions from $\ket{C,j}$ to $\ket{B,i}$, $\ket{B,i}$ to $\ket{D,j}$, and $\ket{D,j}$ to $\ket{B,i}$ respectively. We consider $T=0$ to ease calculations. Under such an assumption, the Fermi-Dirac distribution assumes a Heaviside theta function. We also assume $\gamma_L=\gamma_R=\gamma$. Under such an assumption, in the forward bias, the rates are defined as
\begin{subequations}
\begin{alignat}{8}
R^L_{BC} &= \gamma\left|\bra{C,j}c^\dagger_L\ket{B,i}\right|^2=\gamma\left(\xi\beta+\eta\alpha\right)^2\\
R^R_{BC} &= \gamma\left|\bra{C,j}c^\dagger_R\ket{B,i}\right|^2=\gamma\left(\xi\alpha+\eta\kappa\right)^2\\
R^L_{BD} &= \gamma\left|\bra{D,j}c^\dagger_L\ket{B,i}\right|^2=\gamma\eta^2\\
R^R_{BD} &= \gamma\left|\bra{D,j}c^\dagger_R\ket{B,i}\right|^2=\gamma\xi^2
\end{alignat}
\label{eq:matrix_elements_appendix}
\end{subequations}
and all the other rates are zero.
Since all four $B$ states occur with equal probabilities, the current is given as
\begin{align}
    I=4\mathbb{I}_\mathcal{C}R^L_{BC}P_B + 4\mathbb{I}_\mathcal{D}R^L_{BD}P_B
\end{align}
where $\mathbb{I}_\mathcal{C}\left(\mathbb{I}_\mathcal{D}\right)=1$ if at least one $\mathcal{C}(\mathcal{D})$ transition is accessed within the bias window, and zero otherwise.

We need to use the master \eqref{eq:master_equation} to solve for the probabilities of the states. Let us go case by case.
\subsection{Only one $\mathcal{C}$ transition is accessed}
Let us say one transition $\mathcal{C}_{\zeta_2}$ is accessed. This means that from the four states of the form $B_{\zeta_1}$, only three states of the form $C_{\zeta_1\zeta_2}$ can be accessed via an allowed transition. Solving for the probabilities, we get
\begin{align}
    P_B = \frac{1}{4+3\frac{R^L_{BC}}{R^R_{CB}}}
\end{align}
so that the current \eqref{eq:current} takes the form
\begin{align}
    I_1 = \frac{4R^L_{BC}}{4+3\frac{R^L_{BC}}{R^R_{CB}}} = \frac{4\left(\xi\beta+\eta\alpha\right)^2}{4+3\frac{\left(\xi\beta+\eta\alpha\right)^2}{\left(\xi\alpha+\eta\kappa\right)^2}}\label{eq:app:I1}
\end{align}

\subsection{One pair of $\mathcal{C}$ and $\mathcal{D}$ transition is accessed}
Let us say the transitions $\mathcal{C}_{\zeta_2}$ and $\mathcal{D}_{\zeta_2}$ are accessed. This means that from the four states of the form $B_{\zeta_1}$, only three states of the form $C_{\zeta_1\zeta_2}$ and three states of the form $D_{\zeta_1\zeta_2}$ can be accessed via an allowed transition. Solving for the probabilities, we get
\begin{align}
    P_B = \frac{1}{4+3\frac{R^L_{BC}}{R^R_{CB}}+3\frac{R^L_{BD}}{R^R_{DB}}}
\end{align} 
so that the current \eqref{eq:current} takes the form
\begin{align}
    I_2 = \frac{4R^L_{BC}+4R^L_{BD}}{4+3\frac{R^L_{BC}}{R^R_{CB}}+3\frac{R^L_{BD}}{R^R_{DB}}} = \frac{4\left(\xi\beta+\eta\alpha\right)^2+4\eta^2}{4+3\frac{\left(\xi\beta+\eta\alpha\right)^2}{\left(\xi\alpha+\eta\kappa\right)^2}+3\frac{\eta^2}{\xi^2}}\label{eq:app:I2}
\end{align}

\subsection{Two pairs of $\mathcal{C}$ and $\mathcal{D}$ transitions are accessed}
Now, from the four states of the form $B$, five states in $C$ and five states in $D$ can be accessed via an allowed transition. Solving for the probabilities, we get
\begin{align}
    P_B = \frac{1}{4+5\frac{R^L_{BC}}{R^R_{CB}}+5\frac{R^L_{BD}}{R^R_{DB}}}
\end{align}
so that the current \eqref{eq:current} takes the form
\begin{align}
    I_3 = \frac{4R^L_{BC}+4R^L_{BD}}{4+5\frac{R^L_{BC}}{R^R_{CB}}+5\frac{R^L_{BD}}{R^R_{DB}}} = \frac{4\left(\xi\beta+\eta\alpha\right)^2+4\eta^2}{4+5\frac{\left(\xi\beta+\eta\alpha\right)^2}{\left(\xi\alpha+\eta\kappa\right)^2}+5\frac{\eta^2}{\xi^2}}\label{eq:app:I3}
\end{align}

It is obvious that $I_3<I_2$. Under the conditions chosen in the main text (Sec.~\ref{sec:results}), namely, $\xi\ll\eta$, and $\beta\ll\alpha\approx\kappa$, the denominators of \eqref{eq:app:I2} and \eqref{eq:app:I3} become extremely large, thereby causing a blockade. We thus obtain the result $I_3<I_2\ll I_1$, as is reflected in Fig~\ref{fig:ndr_physics}(b).

\bibliography{main}

\end{document}